\documentclass[twocolumn]{jpsj2}
\usepackage{amsmath, amsfonts, amsthm, amssymb}
\usepackage{graphicx}
\usepackage{bm}
\usepackage{cite}
\usepackage{accents}
\usepackage{enumerate}
\def\runauthor{\textsc{D.~Tahara} and \textsc{M.~Imada}}
\makeatletter
\def\@typeset{}
\makeatother
\def\journal#1#2#3#4{#1 {\bf #2} (#4) #3}
\def\JPSJ{J.\ Phys.\ Soc.\ Jpn.}
\def\PR{Phys.\ Rev.}
\def\PRL{Phys.\ Rev.\ Lett.}
\def\PRB{Phys.\ Rev.\ B}
\def\PRD{Phys.\ Rev.\ D}
\def\RMP{Rev.\ Mod.\ Phys.}
\def\JCP{J.\ Chem.\ Phys.}
\def\etal{{\itshape et al.}}
\renewcommand{\[}{\begin{equation}}
\renewcommand{\]}{\end{equation}}
\newcommand{\bdot}{\bm{\cdot}}
\newcommand{\Ha}{\mathcal{H}}
\newcommand{\mh}{\mathsf{h}}
\newcommand{\mA}{\mathsf{A}}
\newcommand{\mB}{\mathsf{B}}
\newcommand{\mC}{\mathsf{C}}
\newcommand{\mS}{\mathsf{S}}
\newcommand{\mU}{\mathsf{U}}
\newcommand{\mX}{\mathsf{X}}
\newcommand{\sP}{\mathcal{P}}
\newcommand{\sL}{\mathcal{L}}
\newcommand{\sO}{\mathcal{O}}
\newcommand{\Pf}{\mathrm{Pf}\,}
\newcommand{\Tr}{\mathrm{Tr}\,}
\newcommand{\ep}{\varepsilon}
\newcommand{\la}{\langle}
\newcommand{\ra}{\rangle}
\newcommand{\ga}{\alpha}
\newcommand{\gb}{\beta}
\newcommand{\gc}{\gamma}
\newcommand{\gs}{\sigma}
\newcommand{\vk}{{\bm{k}}}

\newcommand{\vr}{{\bm{r}}}

\newcommand{\vQ}{{\bm{Q}}}
\newcommand{\vga}{{\bm{\alpha}}}
\newcommand{\vgc}{{\bm{\gamma}}}
\newcommand{\Ns}{N_{\text{s}}}
\makeatletter
\def\widebar{\accentset{\overline{\mskip12mu}}}
\makeatother
\title{%
Variational Monte Carlo Method Combined with Quantum-Number \\
Projection and Multi-Variable Optimization 
}
\author{%
Daisuke \textsc{Tahara} and Masatoshi \textsc{Imada}$^{1,2}$
}
\inst{%
Department of Physics, University of Tokyo, 7-3-1 Hongo, Bunkyo-ku, Tokyo 113-0033\\
$^1$Department of Applied Physics, University of Tokyo, 7-3-1 Hongo, Bunkyo-ku, Tokyo 113-8656\\
$^2$JST, CREST, Hongo, Bunkyo-ku, Tokyo 113-8656
}
\abst{%
Variational wave functions used in the variational Monte Carlo (VMC) method are extensively improved to overcome the biases coming from the assumed variational form of the wave functions. 
We construct a highly generalized variational form by introducing a large number of variational parameters to the Gutzwiller-Jastrow factor as well as to the one-body part. Moreover, the projection operator to restore the symmetry of the wave function is introduced. 
These improvements enable to treat fluctuations with long-ranged as well as short-ranged correlations. 
A highly generalized wave function is implemented 
by the Pfaffians introduced by Bouchaud \etal, 
together with 
the stochastic reconfiguration method introduced by Sorella for the parameter optimization. 
Our framework offers much higher accuracy 
for strongly correlated electron systems 
than the conventional variational Monte Carlo methods.
}
\kword{%
variational Monte Carlo method, strongly correlated electron systems, Hubbard model, quantum-number projection, stochastic reconfiguration method
}
\begin{document}
\maketitle
%
\section{Introduction}
\label{Sec:Intro}
Strongly correlated electron systems have brought 
many fundamental and challenging 
issues in condensed matter physics \cite{ImadaFujimoriTokura}. 
They are characterized by a competition between the itinerancy of electrons 
favored by the kinetic energy and the localization caused by 
the Coulomb interaction. 
When the latter contribution becomes predominant, the material 
turns 
from metal into the Mott insulator 
at specific electron 
densities \cite{Mott}. 
This metal-insulator transition is called the Mott transition. 
Mott insulators and its related materials show fruitful 
properties such as the high-temperature superconductivity in copper oxides \cite{Bednorz}. 
Such phenomena are certainly beyond the framework of the standard
band theory based on the one-body approximation.
Many-body correlation effects play crucial roles in the strongly correlated systems.
\par
Theoretical routes to investigate these systems are severely 
restricted because of difficulties in treating 
strong correlation effects. 
For this purpose, 
there 
exist several numerical methods, such as 
the exact diagonalization (ED), 
auxiliary-field quantum Monte Carlo (AFQMC) \cite{QMC01,QMC02,AFQMC,FurukawaImada}, 
density matrix renormalization group (DMRG) \cite{DMRG}, 
dynamical mean-field theory (DMFT) \cite{DMFT01,DMFT02}, 
path-integral renormalization group (PIRG) \cite{Kashima,Morita,PIRGx01,PIRGx02,GPIRG,PIRGMizusaki}, 
Gaussian-basis Monte Carlo (GBMC) \cite{GBMC01,GBMCAssaad,GBMCAimi}, 
and 
variational Monte Carlo (VMC) \cite{CeperleyVMC} methods. 
\par
Among them, the VMC method is tractable 
in relatively large system sizes even 
at large amplitude of
interactions and geometrical frustrations. 
However, the bias inherently and inevitably contained in the assumed variational form 
of the wave functions is 
a fundamental drawback in the VMC method. 
Therefore, construction of
highly accurate wave functions is crucially important.
In an interesting region where various phases compete, wave functions which do not 
sufficiently take into account quantum fluctuation effects 
often give even qualitatively wrong results.
\par
The VMC method \cite{CeperleyVMC} offers 
the exact treatment of 
Jastrow-type wave functions \cite{Jastrow} 
within the statistical accuracy. 
The Gutzwiller-Jastrow factors \cite{Jastrow,Gutzwiller}, which are operated to one-body wave functions, enable to take account of 
many-body correlation effects and go beyond mean-field descriptions. 
However, in conventional treatments, the one-body parts are usually 
simply taken as the ground state of the mean-field Hamiltonian where the symmetry is explicitly broken by the mean field. 
The limitations and drawbacks of these conventional variational wave functions are the following:
\begin{enumerate}[(i)]
\item It is hard to describe different competing phases within a single variational form.
\item They do not often satisfy inherent symmetry properties because of the symmetry-broken one-body part.
\item Although the one-body part crucially determines fundamental properties of the variational wave function, the one-body part remains primitive if quantum fluctuation effects are not taken into account.
\end{enumerate}
\par
Recently, numerical techniques to optimize a huge number of 
variational parameters in the VMC framework are 
developed \cite{SorellaSR,SorellaSRH,VMCOptLM02}.
These developments have opened the possibility of overcoming a biased nature of the variational approach and allow us to extend the potential of variational wave function study. 
The biases inherently and inevitably contained in the assumed variational form of the wave functions are aimed to be largely relaxed by a large number of variational parameters, which allow us to treat fluctuations with long-ranged as well as short-ranged correlations. 
One of the successful results in this approach is seen in the electron-state calculations for small molecules \cite{VMCOptLM02,VMCOptNW01,VMCOptLM01,VMCOptNW02,VMCOptEFP}. In these studies, a linear combination of multi-configurational Slater determinants with the Jastrow factor is chosen as a variational wave function. All the parameters, such as linear coefficients, orbitals in Slater determinants, and Jastrow parameters, are optimized by the recently developed energy minimization techniques. These efforts offer a reliable method to obtain quantitatively accurate wave functions in small molecules. 
However, this treatment can not be directly applied to the bulk electron systems, because we have to deal with hundreds of electrons and can not handle multi-configurational Jastrow wave functions within practical computational costs.
In the VMC studies on lattice models, a large number of variational parameters have been introduced to the Jastrow factor by Sorella \cite{SorellaSR}. 
This improvement has opened possibility to describe quantum phase transitions within a single variational wave function \cite{CapelloPRL01}. 
However, biases coming from the one-body part still remain 
because the conventional one-body part corresponds to the mean-field single Slater determinant with only a few variational parameters. This hypothesis strongly influences variational results even though the Gutzwiller-Jastrow factor with many parameters is introduced. 
In order to reduce the conventional biases, the one-body part must be reconstructed by a deliberately examined parameterization.
\par
In this paper, 
we reconstruct and improve the one-body part by introducing many variational parameters with well-thought-out and computationally tractable forms. 
We also introduce 
several symmetry projections in the ground state. 
We demonstrate the efficiency and accuracy 
of our variational framework in which many variational parameters and the symmetry projection allow reducing the biases and bring us quantitatively more accurate wave functions than those in the literature. 
Our goal is to introduce 
a conceptually new scheme for strongly correlated electrons under large quantum fluctuations. This is crucially important in simulating regions near the quantum critical points and regions of competing orders with enhanced fluctuations.
\par
Single-band Hubbard model is suited for benchmark of many-body correlation effects. 
We improve variational wave functions in order to study the ground-state properties of the Hubbard model on a square lattice defined by
\[ \label{eq:FruHub}
  \Ha = \sum_{\vk,\gs} \ep(\vk) c_{\vk\gs}^\dag c_{\vk\gs} + U\sum_{i} n_{i\uparrow} n_{i\downarrow}
,
\]
where
\[
  c_{\vk\gs}^\dag = \frac{1}{\sqrt{\Ns}} \sum_{i} c_{i\gs}^\dag e^{i\vk\bdot\vr_i} \, , \ 
  c_{\vk\gs} = \frac{1}{\sqrt{\Ns}} \sum_{i} c_{i\gs} e^{-i\vk\bdot\vr_i}
\]
are the creation and annihilation operators. The number operator is $n_{i\gs} = c_{i\gs}^\dag c_{i\gs}$. The energy dispersion $\ep(\vk)$ is given by 
\[
  \ep(\vk) = -2t (\cos k_x + \cos k_y) - 4 t' \cos k_x \cos k_y
,
\]
where $t$ ($t'$) is the transfer integral between the Wannier orbitals of nearest-neighbor (next nearest-neighbor) sites.
Model parameters are $t'/t$, $U/t$, and filling $n=N/\Ns$, where $N$ is the total number of electrons and $\Ns$ is the total number of sites. We take $\Ns = L\times L$ sites with 
the boundary condition periodic in $x$ direction and antiperiodic in $y$ direction 
(periodic-antiperiodic boundary condition).
\par
The organization of this paper is as follows. In \S 2, we introduce variational wave functions used in this study. The functional form with a large number of variational parameters and the symmetry projection are described. The optimization method to efficiently handle many variational parameters is explained in \S 3. 
The accuracy of our variational wave functions is benchmarked by comparisons 
with results obtained from unbiased methods in \S 4. 
Section 5 is devoted to summary and discussions.
\section{Variational Wave Functions}
\label{Ch:VWF}
In this section, we introduce 
variational wave functions used in this paper. 
We construct wave functions which bear 
the following properties:
\begin{enumerate}[(i)]
\item Flexibility to describe several different phases by controlling many variational parameters 
of a unified variational form,
\item Capability of treating many-body 
correlations beyond mean-field states,
\item Conservation of symmetry with quantum numbers expected in the ground state.
\end{enumerate}
Recent development in the VMC method allows us to deal with a large number of parameters \cite{SorellaSR,SorellaSRH,VMCOptLM02}. These numerical techniques 
are described in \S \ref{Ch:Opt}. 
We construct wave functions which enhance the capability of removing biases posed on the variational form. This is achieved at least partially by introducing enormous number of parameters.
\subsection{Functional form of variational wave functions}
The general functional form of wave functions in this paper is
\[
  |\psi\ra = \sP \sL | \phi \ra,
\]
where $|\phi\ra$ is a Hartree-Fock-Bogoliubov type wave function called ``one-body part,''
$\sL$ is the quantum-number projector \cite{ManyBody, PIRGMizusaki} controlling symmetries of wave function, and 
$\sP$ is the Gutzwiller-Jastrow factor \cite{Jastrow,Gutzwiller} including many-body correlations.
In order to improve variational wave functions within the sector classified by quantum numbers, 
we only employ $\sP$ that preserves symmetries of $\sL|\phi\ra$.
This means that $\sL$ and $\sP$ are commutable ($\sP\sL=\sL\sP$).
\subsubsection{One-body part}
\label{Ch:VWF-Sec:One-body}
The one-body part usually corresponds to the mean-field Slater determinant with several variational parameters. Though the Gutzwiller-Jastrow factor introduces many-body correlations, this variational hypothesis strongly restricts flexibility of wave functions. 
We reexamine the functional form of the one-body part and introduce as many as 
possible variational parameters in order to improve wave functions.
\par
First, we consider a Hartree-Fock-Bogoliubov type wave function with antiferromagnetic (AF) and superconducting (SC) orders which have been introduced by Giamarchi and Lhuillier \cite{Giamarchi}. Here, we start from a slightly different representation introduced by Himeda and Ogata \cite{HimedaOgata}. The wave function diagonalizes the mean-field Hamiltonian
\begin{align} \notag
  \Ha_{\text{MF}} = &\sum_{\vk\in\text{AFBZ}}
  \biggl[\ep^{a} (\vk) 
   \Bigl( a_{\vk\uparrow}^{\dag} a_{\vk\uparrow} + a_{\vk\downarrow}^{\dag} a_{\vk\downarrow}\Bigr)
\\ \notag
&\qquad\qquad
  + \varDelta_{\text{SC}}^{a}(\vk) \Bigl(a_{\vk\uparrow}^\dag a_{-\vk\downarrow}^\dag + a_{-\vk\downarrow} a_{\vk\uparrow}\Bigr)
\\ \notag
  &\qquad\qquad+\ep^{b} (\vk) 
  \Bigl( b_{\vk\uparrow}^{\dag} b_{\vk\uparrow} + b_{\vk\downarrow}^{\dag} b_{\vk\downarrow}\Bigr)
\\ \notag
  &\qquad\qquad+ \varDelta_{\text{SC}}^{b}(\vk) \Bigl(b_{\vk\uparrow}^\dag b_{-\vk\downarrow}^\dag + b_{-\vk\downarrow} b_{\vk\uparrow}\Bigr)
  \biggr]
\\&-
  \mu \sum_{i,\gs} c_{i\gs}^\dag c_{i\gs}
,
\label{eq:VWF-HamMF}
\end{align}
where AFBZ denotes the folded AF Brillouin zone, $\varDelta_{\text{SC}}^{a}(\vk)$ and $\varDelta_{\text{SC}}^{b}(\vk)$ are SC order parameters, and $\mu$ is the chemical potential. 
The energy dispersion of AF bands $\ep^{a}(\vk)$ and $\ep^{b}(\vk)$ are defined as
\[
  \left\{
  \begin{array}{l}
    \ep^{a}(\vk) = \xi_2(\vk) - \sqrt{\xi_1(\vk)^2+\varDelta_{\text{AF}}^2}\\[+5pt]
    \ep^{b}(\vk) = \xi_2(\vk) + \sqrt{\xi_1(\vk)^2+\varDelta_{\text{AF}}^2}
  \end{array}
  \right.
  \Bigl( \vk \in \text{AFBZ} \Bigr)
\]
with $\xi_1(\vk) = (\ep(\vk)-\ep(\vk+\vQ))/2$ and $\xi_2(\vk) = (\ep(\vk)+\ep(\vk+\vQ))/2$, 
where $\varDelta_{\text{AF}}$ is the AF order parameter and the vector $\vQ$ corresponds to the AF order ($\vQ=(\pi,\pi)$).
The operators $\{a_{\vk\gs}^\dag, b_{\vk\gs}^\dag\}$ ($\{a_{\vk\gs}, b_{\vk\gs}\}$) are creation (annihilation) operators for the AF quasiparticles related to the electron operator $c_{\vk\gs}^\dag$ ($c_{\vk\gs}$) through the following unitary transformation
\[ \label{eq:VWF-akbk}
  \left\{
  \begin{array}{l}
    a_{\vk\gs}^\dag = u_{\vk} c_{\vk\gs}^\dag + \gs v_{\vk} c_{\vk+\vQ\gs}^\dag \\[+5pt]
    b_{\vk\gs}^\dag = -\gs v_{\vk}c_{\vk\gs}^\dag + u_{\vk} c_{\vk+\vQ\gs}^\dag
  \end{array}
  \right.
  \Bigl( \vk \in \text{AFBZ} \Bigr)
\]
with
\[ \label{eq:VWF-ukvk}
  u_{\vk} (v_{\vk}) 
  = \Biggl[ \dfrac{1}{2} \biggl( 1-(+)\dfrac{\xi_1(\vk)}{\sqrt{\xi_1(\vk)^2+\varDelta_{\text{AF}}^2}} \biggr) \Biggr]^{1/2}
.
\]
\par
The wave function with $N$ particles extracted from the eigenfunction of $\Ha_{\text{MF}}$ is written as
\[ \label{eq:VWF-AFSC}
  | \phi_{\text{AF+SC}} \ra \!
=\!
  \Biggl[ \sum_{\vk\in\text{AFBZ}}
    \!\!\!\!\Bigl(
    \varphi^{a}(\vk) a_{\vk\uparrow}^\dag a_{-\vk\downarrow}^\dag
+
    \varphi^{b}(\vk) b_{\vk\uparrow}^\dag b_{-\vk\downarrow}^\dag
    \Bigr)
  \Biggr]^{N/2} \!\!\!\!\!| 0 \ra
\]
with
\[ \label{eq:VWF-varphiab}
\left\{
\begin{array}{@{}l@{}} \displaystyle
  \varphi^{a}(\vk)
  = \frac{\varDelta_{\text{SC}}^{a}(\vk)}{ (\ep^a(\vk) - \mu) + \sqrt{
    (\ep^a(\vk) - \mu)^2 + [\varDelta_{\text{SC}}^{a}(\vk)]^2
  } }
\\[+15pt] \displaystyle
  \varphi^{b}(\vk)
  = \frac{\varDelta_{\text{SC}}^{b}(\vk)}{ (\ep^b(\vk) - \mu) + \sqrt{
    (\ep^b(\vk) - \mu)^2 + [\varDelta_{\text{SC}}^{b}(\vk)]^2
  } }
\end{array}
\right.
.
\]
In the limit $\varDelta_{\text{AF}}\to 0$, the operators $a_{\vk\gs}^\dag$ and $b_{\vk\gs}^\dag$ are reduced to $c_{\vk\gs}^\dag$ and $c_{\vk+\vQ\gs}^\dag$, respectively. The wave function $|\phi_{\text{AF+SC}}\ra$ (eq. (\ref{eq:VWF-AFSC})) becomes the conventional BCS wave function. On the other hand, in the limit $\varDelta_{\text{SC}}^{a}(\vk) (\varDelta_{\text{SC}}^{b}(\vk)) \to 0$, $\varphi^{a}(\vk) (\varphi^{b}(\vk))$ goes to zero if $\ep^{a}(\vk) (\ep^{b}(\vk)) > \mu$ and otherwise diverges. Thus, the AF quasiparticles are only filled 
below 
the chemical potential and $|\phi_{\text{AF+SC}}\ra$ is reduced to the normal AF mean-field wave function.
\par
In VMC studies on lattice systems, several variational parameters are considered to 
improve the one-body part. For example, $\varDelta_{\text{AF}}$ and $\varDelta_{\text{SC}}^{a(b)}(\vk)$ are adopted to describe the magnetism and superconductivity, respectively. The chemical potential $\mu$ and band renormalization effects \cite{HimedaOgata} improve the accuracy.
These variational parameters allow optimizations of $u_{\vk}$, $v_{\vk}$, $\varphi^{a}(\vk)$, and $\varphi^{b}(\vk)$ to a certain but restricted extent in eqs. (\ref{eq:VWF-ukvk}) and (\ref{eq:VWF-varphiab}).
\par
Now, instead of taking $\varDelta_{\text{AF}}$, $\varDelta_{\text{SC}}^{a(b)}(\vk)$, and $\mu$ as variational parameters, we take parameters $u_{\vk}$, $v_{\vk}$, $\varphi^{a}(\vk)$, $\varphi^{b}(\vk)$ independently 
for each $\vk\in\text{AFBZ}$ under the conditions:
\begin{gather}
  u_{\vk}^2+v_{\vk}^2=1 \, , \ 
  u_{-\vk} = u_{\vk} \, , \ 
  v_{-\vk} = v_{\vk} \, , \\
  \varphi^{a}(-\vk) = \varphi^{a}(\vk) \, , \ 
  \varphi^{b}(-\vk) = \varphi^{b}(\vk).
\end{gather}
From eqs. (\ref{eq:VWF-akbk}) and (\ref{eq:VWF-AFSC}), $|\phi_{\text{AF}+\text{SC}}\ra$ can be written with $c_{\vk\gs}^{\dag}$:
\begin{align} \notag
  |\phi_{\text{AF}+\text{SC}}\ra =&
  \Biggl[ \sum_{\vk\in\text{AFBZ}}
    \biggl\{
    \Bigl( u_{\vk}^2 \varphi^{a}(\vk) - v_{\vk}^2 \varphi^{b}(\vk) \Bigr)
    c_{\vk\uparrow}^\dag c_{-\vk\downarrow}^\dag
\\ \notag
&+
    \Bigl( -v_{\vk}^2 \varphi^{a}(\vk) + u_{\vk}^2 \varphi^{b}(\vk) \Bigr)
    c_{\vk+\vQ\uparrow}^\dag c_{-\vk-\vQ\downarrow}^\dag
\\ \notag
&+
    \Bigl( \varphi^{a}(\vk) + \varphi^{b}(\vk) \Bigr) u_{\vk}v_{\vk}
\\[-10pt] \label{eq:VWF-029}
&\qquad  \times  \Bigl( 
        c_{\vk+\vQ\uparrow}^\dag c_{-\vk\downarrow}^\dag
      - c_{\vk\uparrow}^\dag c_{-\vk-\vQ\downarrow}^\dag
    \Bigr)
    \biggr\}
  \Biggr]^{N/2} \!\!| 0 \ra
.
\end{align}
Then, we transform $u_{\vk}$, $v_{\vk}$, $\varphi^{a}(\vk)$, $\varphi^{b}(\vk)$ to
\[
  \left\{
\begin{array}{l}
  u_{\vk} = \cos \theta_{\vk} \, , \ v_{\vk} = \sin \theta_{\vk} \\[+5pt]
  A(\vk)  = \cos^2 \theta_{\vk} \varphi^{a}(\vk) - \sin^2 \theta_{\vk} \varphi^{b}(\vk)  \\[+5pt]
  B(\vk)  = -\sin^2 \theta_{\vk} \varphi^{a}(\vk) + \cos^2 \theta_{\vk} \varphi^{b}(\vk) 
\end{array}
  \right.
.
\]
The coefficient of the third term in eq. (\ref{eq:VWF-029}) is rewritten as
\[
  \Bigl( \varphi^{a}(\vk) + \varphi^{b}(\vk) \Bigr) u_{\vk}v_{\vk} 
=
  \Bigl( A(\vk) + B(\vk) \Bigr) \tan\theta_{\vk}
=
  C(\vk)
.
\]
As a result, variational parameters $u_{\vk}$, $v_{\vk}$, $\varphi^{a}(\vk)$, $\varphi^{b}(\vk)$ are mapped to new parameters $A(\vk)$, $B(\vk)$, $C(\vk)$. 
The parameter $A(\vk)$ ($B(\vk)$) corresponds to singlet pairings in (out) the AFBZ. 
Finally, with the definitions $\varphi^{(1)}(\vk)=A(\vk)$, $\varphi^{(1)}(\vk+\vQ)=B(\vk)$, and $\varphi^{(2)}(\vk)=C(\vk)$ for $\vk \in \text{AFBZ}$, we obtain the wave function
\begin{align} \notag
  |\phi_{\text{pair}}\ra &= 
  \Biggl[ \sum_{\vk\in\text{BZ}}
    \varphi^{(1)}(\vk)
    c_{\vk\uparrow}^\dag c_{-\vk\downarrow}^\dag
\\ \label{eq:VWF-one-bodyfull}
  &+\!\!\!
    \sum_{\vk\in\text{AFBZ}}
\!\!\!
    \varphi^{(2)}(\vk)
    \Bigl( 
        c_{\vk+\vQ\uparrow}^\dag c_{-\vk\downarrow}^\dag
      - c_{\vk\uparrow}^\dag c_{-\vk-\vQ\downarrow}^\dag
    \Bigr)
  \Biggr]^{N/2} \!\!| 0 \ra
\end{align}
with the conditions
\[
  \varphi^{(1)}(-\vk) = \varphi^{(1)}(\vk) \, , \ 
  \varphi^{(2)}(-\vk) = \varphi^{(2)}(\vk)
.
\]
The AF mean-field state is realized by using the second term. 
Dealing with $\varphi^{(1)}(\vk)$ and $\varphi^{(2)}(\vk)$ directly as $\vk$-dependent variational parameters allows us to express various states such as paramagnetic metals, antiferromagnetically ordered states, and superconducting states with any gap function within a single framework of $|\phi_{\text{pair}}\ra$.
Moreover, since the number of the variational parameters increases scaled by the system size, it allows taking account of fluctuation effects with short-ranged correlations.
In this paper, we call $|\phi_{\text{pair}}\ra$ a ``generalized pairing function'' and $\varphi^{(1)}(\vk)$, $\varphi^{(2)}(\vk)$ are called ``pair orbitals.''
Introducing all the possible ordered vectors $\vQ$ would further 
generalize $|\phi_{\text{pair}}\ra$. 
However, this extension 
substantially increases the number of variational parameters 
and computational costs ($\sim\sO(N)$). Therefore, we take 
one physically plausible $\vQ$ in this study.
\par
By using the $\vk$-dependent parameters $\varphi^{(1)}(\vk)$ and the Gutzwiller factor $\sP_{\text{G}}^{\infty}$ (eq. (\ref{eq:VWF-GutzInf})), our variational wave function 
can also represent 
the resonating valence bond (RVB) basis \cite{LDA}, which 
is known to offer 
highly accurate variational wave functions in spin systems. 
We note that the RVB basis can represent the state with spin correlations decaying with arbitrary power laws for increasing distance.
The relation between $\sP_{\text{G}}^{\infty} | \phi_{\text{pair}}\ra$ and the RVB basis is discussed in Appendix \ref{App:RVB}.
\par
In quantum chemistry, Casula \etal{} have introduced a similar wave function called an antisymmetrized geminal power \cite{CasulaJCP}. Since the singlet pairs are only included in this wave function, it is not connected to the AF mean-field wave function. Our extension offers a clear representation to include the singlet pairing wave functions and the AF mean-field wave functions. 
\par
For actual numerical calculations, we rewrite $|\phi_{\text{pair}}\ra$ in a real space representation:
\[
  |\phi_{\text{pair}}\ra = 
  \Biggl[ \sum_{i,j=1}^{\Ns}
    f_{ij} 
    c_{i\uparrow}^\dag c_{j\downarrow}^\dag 
  \Biggr]^{N/2} | 0 \ra
\]
with
\begin{align} \notag
  f_{ij} =& \frac{1}{\Ns} 
  \sum_{\vk\in\text{BZ}} \varphi^{(1)}(\vk) e^{i\vk\bdot(\vr_i-\vr_j)}
 \\ \label{eq:VWF-one-bodyfull-real-space}&+
  \frac{1}{\Ns} 
  \sum_{\vk\in\text{AFBZ}} \varphi^{(2)}(\vk) e^{i\vk\bdot(\vr_i-\vr_j)} \Bigl(e^{i\vQ\bdot\vr_i}-e^{-i\vQ\bdot\vr_j}\Bigr)
.
\end{align}
Here, one of the parameters $\{\varphi^{(1)}(\vk),\varphi^{(2)}(\vk)\}$ is 
not independent because of the normalization of the wave function.
\subsubsection{Gutzwiller-Jastrow factors}
\label{Ch:VWF-Sec:GutzJast}
In the variational study, the Gutzwiller-Jastrow type wave functions \cite{Jastrow,Gutzwiller} are often used to 
take account of 
many-body correlations. The Gutzwiller-Jastrow correlation factor $\sP$ is operated to the one-body wave function $|\phi\ra$, namely as $\sP|\phi\ra$. Since $|\phi\ra$ is usually represented in the $\vk$-space configuration, the factor $\sP$, constructed with many-body operator 
in the real space configuration, introduces compromise of real space and $\vk$-space representations 
into one wave function. 
Because of this uncommutable nature, this factor allows us to go beyond the variational framework of a single Slater determinant and a linear combination of many Slater determinants are generated after the operation of $\sP$, which is crucial in representing strong correlation effects. 
In this paper, we adopt three many-body operators $\sP_{\text{G}}$, $\sP_{\text{d-h}}^{\text{ex.}}$, and $\sP_{\text{J}}$, which are called the Gutzwiller factor, the doublon-holon correlation factor, and the Jastrow factor, respectively.
\par
Gutzwiller has introduced a basic and efficient correlation factor $\sP_{\text{G}}$ \cite{Gutzwiller}, which gives different weights 
to the wave function depending on the rate of double occupancy:
\[
  \sP_{\text{G}} = \exp \biggl[
    -g\sum_{i} n_{i\uparrow} n_{i\downarrow}
  \biggr]
=
  \prod_{i} \Bigl[ 1 - (1-e^{-g}) n_{i\uparrow} n_{i\downarrow} \Bigr]
,
\]
where $g$ is a variational parameter. In the limit $g\to\infty$, $\sP_{\text{G}}$ fully projects out the configurations with finite double occupancy as 
\[ \label{eq:VWF-GutzInf}
  \sP_{\text{G}}^{\infty} = \prod_{i} \Bigl[ 1 -  n_{i\uparrow} n_{i\downarrow} \Bigr]
.
\]
$\sP_{\text{G}}^{\infty}$ is used for the Heisenberg model and the $t$-$J$ model. 
In the Hubbard model with finite $U/t$, the double occupancy is nonzero 
even in the insulating state. Thus we deal with $\sP_{\text{G}}$ at finite $g$.
\par
In order to 
take account of 
many-body effects beyond the Gutzwiller factor, the doublon-holon correlation factor \cite{Kaplan, YokoyamaShiba3} is implemented in the wave function. 
This factor comes from the idea that a doublon (doubly occupied site) 
and a holon (empty site) are bound in the insulator for large $U/t$ \cite{ImadaFujimoriTokura}.
The short-ranged correlation factor with many-body operators has a form
\[ \label{eq:d-h01}
  \sP_{\text{d-h}} = \exp \biggl[
    -\ga_1 \sum_{i} \xi_{i(0)}^{(1)} - \ga_2 \sum_{i} \xi_{i(0)}^{(2)}
  \biggr]
,
\]
where $\ga_1$ and $\ga_2$ are variational parameters. Here, 
$\xi_{i(m)}^{(\ell)}$ is a many-body operator which is diagonal in the real space representations and may be given by 
\[
  \xi_{i(m)}^{(\ell)}
=
  \left\{
\begin{array}{@{\ }c@{\ \ }l@{}}
  1 & \left[\mbox{\parbox{4.5cm}{if a doublon (holon) exists at the site $i$ and $m$ holons (doublons) surround at the $\ell$-th nearest neighbor}}\right]
\\[+7mm]
  0 & \,[\,\text{otherwise}\,]
\end{array}
  \right.
.
\]
For example, $\xi_{i(0)}^{(1)}$ is written by
\[
  \xi_{i(0)}^{(1)} = d_i \prod_{\tau}^{\text{n.n.}} (1-h_{i+\tau}) + h_i \prod_{\tau}^{\text{n.n.}} (1-d_{i+\tau})
,
\]
where the product $\prod_{\tau}^{\text{n.n.}}$ runs over nearest-neighbor sites, and $d_i=n_{i\uparrow}n_{i\downarrow}$ and $h_i = (1-n_{i\uparrow})(1-n_{i\downarrow})$ are doublon and holon 
operators, respectively. 
The doublon-holon correlation factor $\sP_{\text{d-h}}$ given by eq. (\ref{eq:d-h01})
or by slightly different forms has been adopted in several VMC studies \cite{Kaplan,YokoyamaShiba3,Liu,WataVMC01,YokoVMCdiag}. 
Recently, there is a proposal to extend eq. (\ref{eq:d-h01}) by introducing 
many $\xi_{i(m)}^{(\ell)}$ \cite{KobayashiYokoyama}. We take $\sP_{\text{d-h}}^{\text{ex.}}$ as
\[
  \sP_{\text{d-h}}^{\text{ex.}} = \exp\biggl[
   - \sum_{m=0}^{2}\sum_{\ell=1,2} \ga_{(m)}^{(\ell)} \sum_{i} \xi_{i(m)}^{(\ell)}
  \biggr]
,
\]
where $\ga_{(m)}^{(\ell)}$ are variational parameters. 
It is in principle possible to include operators with $m=3,4, \ldots$, 
but contributions of higher $m$ parts have turned out to be negligible while have induced instabilities in our optimization procedure. 
Therefore we confine ourselves to $m$ up to $2$.
\par
\begin{figure}[t]
\centering
\includegraphics[width=0.45\textwidth]{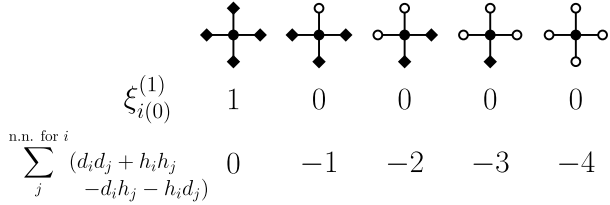}
\caption{%
Weights of the many-body operators in $\sP_{\text{d-h}}$ (upper row) and $\sP_{\text{J}}$ (lower row) for some configurations. 
Filled circles, open circles, and filled diamonds denote doublon sites, holon sites, and single occupied sites, respectively.
}
\label{fig:VWF-dhjast}
\end{figure}

Jastrow has introduced a long-ranged correlation factor for continuum systems \cite{Jastrow}. This factor takes into account correlation effects through two-body operators. In the Hubbard model at quarter filling, Yokoyama and Shiba have discussed the effects of the Jastrow-type correlation factor \cite{YokoyamaShiba3}. Recently, Capello \etal {} have claimed a necessity of this factor to describe the Mott transition \cite{CapelloPRL01}.The Jastrow factor $\sP_{\text{J}}$ in lattice models has the following form:
\[
  \sP_{\text{J}} = \exp \biggl[
    - \frac{1}{2} \sum_{i\neq j} v_{ij} n_{i} n_{j}
  \biggr] 
\]
with two-body terms, 
where $n_{i}=n_{i\uparrow}+n_{i\downarrow}$ is a density operator and $v_{ij}=v(\vr_i-\vr_j)$ are variational parameters depending on the displacement $\vr_i-\vr_j$.
The on-site Jastrow factor is equivalent to the Gutzwiller factor except for a constant factor:
\begin{align} \notag
  \sP_{\text{J}}^{\text{on-site}} &= 
  \exp \biggl[
    -\frac{1}{2} \sum_{i} v_{ii} n_{i} n_{i}
  \biggr]
\\
&=
  \underbrace{\exp \biggl[-\frac{1}{2}v(0) N\biggr]}_{\displaystyle \text{const.}}
  \underbrace{
  \exp \biggl[-
    v(0)\sum_{i} v_{ii} n_{i\uparrow} n_{i\downarrow}
  \biggr]
  }_{\displaystyle = \sP_{\text{G}}}
.
\end{align}
From the viewpoint of doublon-holon correlations, $\sP_{\text{J}}$ can be rewritten as doublon-doublon (holon-holon) repulsive and doublon-holon attractive operators:
\begin{gather} \notag
  \sum_{i\neq j} v_{ij} n_{i} n_{j}
=
  \sum_{i\neq j} v_{ij} (d_id_j+h_ih_j-d_ih_j-h_id_j) + C,
\\
  \sP_{\text{J}} \propto
  \exp\biggl[
    -\frac{1}{2} \sum_{i\neq j} v_{ij} (d_id_j+h_ih_j-d_ih_j-h_id_j)
  \biggr]
,
\end{gather}
where $C$ is some constant. 
We remark the difference between the doublon-holon correlation factor and 
the Jastrow factor: For example, the two many-body operators 
give different weights to the configurations as 
shown in Fig. \ref{fig:VWF-dhjast}.
\subsubsection{Quantum-number projection}
In general, quantum many-body systems have several symmetries related to the Hamiltonian such as translational symmetry, point group symmetry of lattice, $U(1)$ gauge symmetry, and $SU(2)$ spin-rotational symmetry. 
While symmetry breaking occurs in the thermodynamic limit, these symmetries must be preserved in finite many-body systems.
\par
Variational wave functions constructed from one-body parts and the Gutzwiller-Jastrow factors do not often satisfy inherent symmetry properties, because the Hartree-Fock-Bogoliubov type one-body part comes from symmetry broken mean-field treatment. 
Even in the generalized pairing wave function $|\phi_{\text{pair}}\ra$, the spin-rotational symmetry is broken by the orbital $\varphi^{(2)}(\vk)$ which enables to include the mean-field AF state.
\par
The quantum-number projection technique \cite{ManyBody} enables to control symmetries of wave function. This technique has been used successfully in the PIRG method \cite{PIRGMizusaki} and the GBMC method \cite{GBMCAssaad, GBMCAimi}. 
By using the quantum-number projection together with the Gutzwiller-Jastrow factor, one can construct variational wave functions with controlled symmetries and many-body correlations.
The quantum-number projection operator $\sL$ is constructed by superposing transformation operators $T^{(n)}$ with weights $w_{n}$:
\[ \label{eq:QPgeneral}
  \sL |\phi\ra = \sum_{n} w_{n} T^{(n)} |\phi\ra = \sum_{n} w_{n} |\phi^{(n)}\ra
,
\]
where $|\phi\ra$ and $|\phi^{(n)}\ra$ are the original one-body part and the transformed one-body parts, 
respectively. When $\sL$ restores some continuous symmetry, the summation $\sum_{n}$ is replaced by the integration over some continuous variable.
\par
The $SU(2)$ spin-rotational symmetry is restored by superposing wave functions rotated in the spin space. 
The spin projection operator $\sL^{S}$ which filters out $S^z=0$ component of $|\phi\ra$ and generates a state with total spin $S$ and $S^z=0$ has a form
\[
  \sL^S = \frac{2S+1}{8\pi^2} \int d\varOmega \, P_{S}(\cos \beta) R(\varOmega)
,
\]
where $\varOmega=(\ga,\gb,\gc)$ is the Euler angle and the integration is performed over whole range of $\varOmega$. The weight $P_{S}(\cos\beta)$ is the $S$-th Legendre polynomial. The rotational operator $R(\varOmega)$ is defined as
\[
  R(\varOmega) = R^z(\ga) R^y(\gb) R^z(\gc) = e^{i\ga S^z} e^{i\gb S^y} e^{i\gc S^z}
,
\]
where $S^y$ and $S^z$ are total spin operators of $y$ and $z$ directions, respectively.
\par
Now we consider operating $\sL^S$ to the one-body part $|\phi\ra$ which has the form
\[ \label{eq:VWFSP03}
  |\phi\ra = 
  \Biggl[ \sum_{i,j=1}^{\Ns} \sum_{\gs,\gs'=\uparrow,\downarrow}
    f_{ij}^{\gs\gs'} 
    c_{i\gs}^\dag c_{j\gs'}^\dag 
  \Biggr]^{N/2} | 0 \ra
.
\]
The rotated wave function $R(\varOmega)|\phi\ra$ is represented by the same form as eq. (\ref{eq:VWFSP03}) with rotated creation operator $c_{i\gs}^\dag(\varOmega)$:
\[ 
  R(\varOmega)|\phi\ra = 
  \Biggl[ \sum_{i,j=1}^{\Ns} \sum_{\gs,\gs'=\uparrow,\downarrow}
    f_{ij}^{\gs\gs'} 
    c_{i\gs}^\dag(\varOmega) c_{j\gs'}^\dag(\varOmega)
  \Biggr]^{N/2} | 0 \ra
.
\]
The rotated creation operator $c_{i\gs}^\dag(\varOmega)$ is quantized along an axis rotated from $z$ direction:
\[ \label{eq:VWFSP05}
  \left(
  \begin{array}{@{\,}c@{\,}}
    c_{i\uparrow}^\dag (\varOmega) \\
    c_{i\downarrow}^\dag (\varOmega)
  \end{array}
  \right)
= \mathsf{R}^{z}(\ga) \mathsf{R}^{y}(\gb) \mathsf{R}^{z}(\gc)
  \left(
  \begin{array}{@{\,}c@{\,}}
    c_{i\uparrow}^\dag \\
    c_{i\downarrow}^\dag
  \end{array}
  \right)
,
\]
with
\begin{align} \label{eq:VWFSP051}
  \mathsf{R}^{z}(\theta) &= 
  \left(
  \begin{array}{@{\,}cc@{\,}}
    e^{i\theta/2} & 0 \\
    0 & e^{-i\theta/2}
  \end{array}
  \right),
\\ \label{eq:VWFSP052}
  \mathsf{R}^{y}(\theta) &= 
  \left(
  \begin{array}{@{\,}cc@{\,}}
     \cos (\theta/2) & \sin (\theta/2) \\
    -\sin (\theta/2) & \cos (\theta/2) 
  \end{array}
  \right)
.
\end{align}
Thus, $\sL^S|\phi\ra$ has a form
\begin{align} \notag
  \sL^S|\phi\ra &= \frac{2S+1}{8\pi^2} \int d\varOmega \, P_{S}(\cos\beta) 
\\ \notag & \qquad\qquad \times
  \Biggl[ \sum_{i,j,\gs,\gs}
    f_{ij}^{\gs\gs'} 
    c_{i\gs}^\dag(\varOmega) c_{j\gs'}^\dag(\varOmega)
  \Biggr]^{N/2} | 0 \ra
\\ \notag
&=
  \frac{2S+1}{8\pi^2}
  \int_{0}^{2\pi} \!\!\!\!\!d\ga
  \int_{0}^{\pi} \!\!\!\!d\gb
  \int_{0}^{2\pi} \!\!\!\!\!d\gc
  \, \sin\beta \, P_{S}(\cos\beta) 
\\ \label{eq:VWFSP06} & \qquad\qquad \times  \Biggl[ \sum_{i,j,\gs,\gs}
    \widetilde{f}_{ij}^{\gs\gs'} (\varOmega)
    c_{i\gs}^\dag c_{j\gs'}^\dag
  \Biggr]^{N/2} | 0 \ra
,
\end{align}
where $\widetilde{f}_{ij}^{\gs\gs'} (\varOmega)$ is transformed from $f_{ij}^{\gs\gs'}$ by using eqs. (\ref{eq:VWFSP05}), (\ref{eq:VWFSP051}), and (\ref{eq:VWFSP052}). 
\par
The one-body part in this study introduced in \S \ref{Ch:VWF-Sec:One-body} contains 
only $S^z=0$ component $|\phi\ra = [\sum_{ij} f_{ij} c_{i\uparrow}^\dag c_{j\downarrow}^\dag]^{N/2}|0\ra$. Then, the integration over $\gc$ can be omitted and $\sL^S|\phi\ra$ is written as 
\begin{align} \notag
  \sL^S|\phi\ra &= 
  \frac{2S+1}{4\pi}
  \int_{0}^{2\pi} \!\!\!\!\!d\ga
  \int_{0}^{\pi} \!\!\!\!d\gb
  \, \sin\gb\,P_{S}(\cos \beta) 
\\ \label{eq:VWFSP07}
& \qquad\qquad \qquad \times R^z(\ga) R^y(\gb) | \phi\ra
.
\end{align}
In order to omit the integration over $\ga$, we rewrite eq. (\ref{eq:VWFSP07}) to a more convenient form for VMC calculations. 
VMC is performed by sampling of a complete set of real space configurations $\{|x\ra\}$ with which $|\psi\ra$ can be expanded:
\[
  |\psi\ra = \sum_x | x \ra \la x | \psi\ra
.
\]
Since the integration over $\ga$ filters out 
$S^z\neq0$
component, we can chose $\{|x\ra\}$ with $S^z=0$ condition and omit this integration:
\begin{align} \notag
  \sL^S|\phi\ra &= \sum_{x}^{S^z=0} | x\ra\la x|\sL^S|\phi\ra
\\&=
  \sum_{x}^{S^z=0} | x\ra 
  \frac{2S+1}{2}
  \int_{0}^{\pi} \!\!\!\!d\gb
  \, \sin\beta\, P_{S}(\cos\beta) 
  \la x|R^y(\gb)|\phi\ra
.
\end{align}
When the Gutzwiller-Jastrow factor is operated to $\sL^S|\phi\ra$, a similar formula can be obtained as
\[ 
  |\psi\ra = \sP\sL^S|\phi\ra = \sum_{x}^{S^z=0} | x\ra\la x|\sP\sL^S|\phi\ra
\]
with
\begin{align} \notag
  \la x|\sP\sL^S|\phi\ra = 
  P(x)
  \frac{2S+1}{2}
  \int_{0}^{\pi} \!\!\!\!d\gb
  \, &\sin\beta\, P_{S}(\cos\beta) 
\\ \label{eq:VWF-SpinProj}
& \times
  \la x|R^y(\gb)|\phi\ra
.
\end{align}
The integration over $\beta$ is evaluated efficiently by the Gauss-Legendre quadrature in actual numerical calculations \cite{NumRec}. 
Typically, for $S=0$ of the half-filled electron system in $L=4$ and $L=14$ lattices, we need $10$ and $20$ mesh points, respectively.
\par
\begin{figure}[t]
\centering
\includegraphics[width=0.40\textwidth]{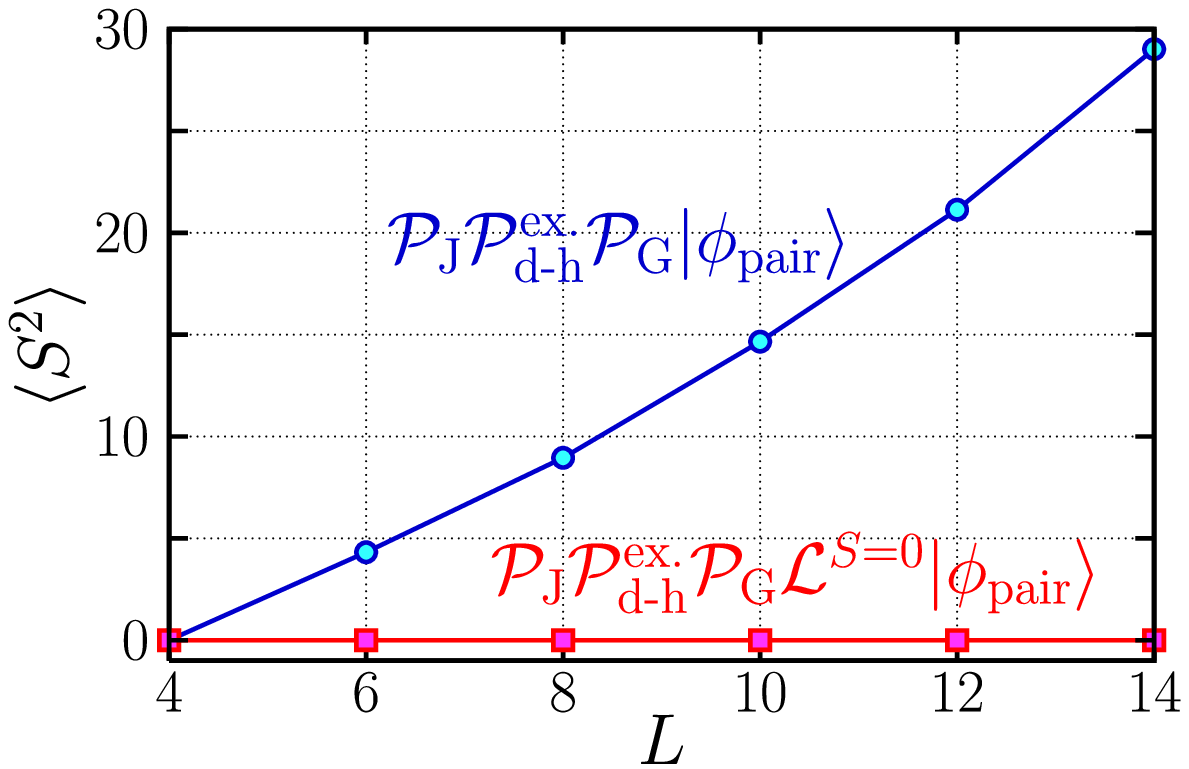}
\caption{%
(Color online) Total spin $\la S^2\ra$ as a function of the system size $L$ at $t'/t=0$, $U/t=4$, $n=1$. 
Error bars are comparable to the symbol size.
}
\label{fig:VWF-SpinProjSS}
\vspace*{5mm}
\centering
\includegraphics[width=0.45\textwidth]{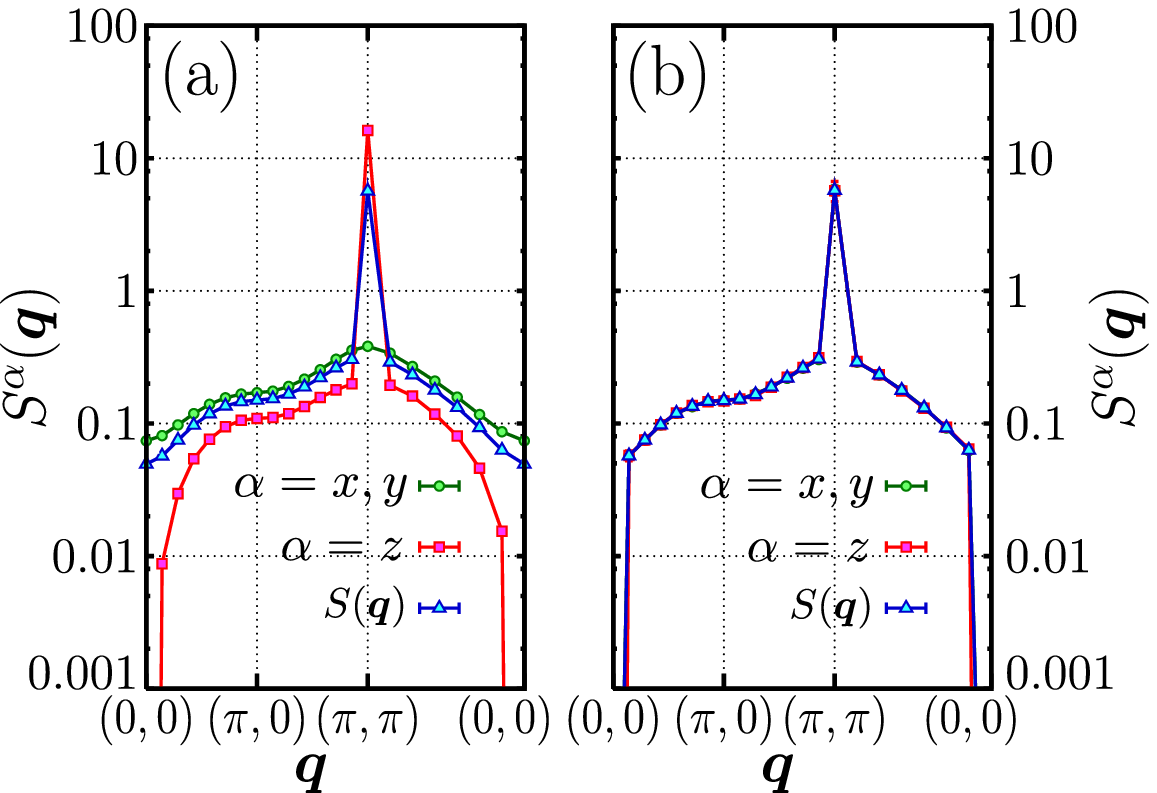}
\caption{%
(Color online) Each element of the spin structure factor obtained by (a) $\mathcal{P}_{\text{J}}\mathcal{P}_{\text{d-h}}^{\text{ex.}} \mathcal{P}_{\text{G}} | \phi_{\text{pair}} \rangle$ and (b) $\mathcal{P}_{\text{J}}\mathcal{P}_{\text{d-h}}^{\text{ex.}} \mathcal{P}_{\text{G}} \mathcal{L}^{S=0}| \phi_{\text{pair}} \rangle$ for $t'/t=0$, $U/t=4$, $n=1$, and $L=14$. 
Error bars are comparable to the symbol size. 
In (b), all the three cases are on the same curve.
}
\label{fig:VWF-SpinProjSq}
\end{figure}
Figure \ref{fig:VWF-SpinProjSS} shows 
the actual VMC simulation results on 
the total spin $\la S^2\ra$ defined as
\[
  \la S^2\ra = \sum_{i,j} \la \bm{S}_{i} \bdot \bm{S}_{j} \ra
.
\]
The total spin 
$\la S^2\ra$ grows as the system size 
increases if the projection is not imposed.
On the other hand, spin projected wave function strictly keeps the $S=0$ state, of course. 
Figure \ref{fig:VWF-SpinProjSq} shows each element of the spin structure factor
\begin{align}
  S^{\ga}(\bm{q}) &= \frac{1}{\Ns} \sum_{i,j} \la S_{i}^{\ga} S_{j}^{\ga}\ra e^{i\bm{q}\bdot(\vr_i-\vr_j)}
 \ (\ga=x,y,z) \, ,
\\
  S(\bm{q}) &= \frac{1}{3} \Bigl( S^x(\bm{q})+S^y(\bm{q})+S^z(\bm{q}) \Bigr)
.
\end{align}
The projected wave function recovers the symmetric property of $S^{\ga}(\bm{q})$. 
\par
Symmetry breaking of the spin part in $|\phi_{\text{pair}}\ra$ is caused by the components of $\varphi^{(2)}(\vk)$, because $\varphi^{(2)}(\vk)$ makes singlet and triplet pairs in $|\phi_{\text{pair}}\ra$. 
There are two ways to restore the spin rotational symmetry. 
The first way is restricting the pair orbitals only to $\varphi^{(1)}(\vk)$ and setting $\varphi^{(2)}(\vk)=0$. Then $|\phi_{\text{pair}}\ra$ has only singlet pairs and the spin projection can be omitted. 
The second way is dealing with all orbitals $\varphi^{(1)}(\vk)$, $\varphi^{(2)}(\vk)$ 
which requires an additional spin projection.
\par
Table \ref{Tbl:VWF-Restrict} shows the variational energy calculated by the above two ways. The wave function on the latter way is better than one on the former. The combination of spin projection and symmetry breaking pair orbitals provides much accurate wave functions.
\begin{table}[t]
\caption{Variational energy of wave function with (a) only singlet pairs and (b) symmetry breaking orbitals and spin projection for 
$L=4$, $t'/t=0$, $U/t=5$, $n=1$. The Gutzwiller-Jastrow factor is $\sP=\sP_{\text{J}}\sP_{\text{d-h}}^{\text{ex.}}\sP_{\text{G}}$. The number in parentheses is the statistical error in the last digits.}
\label{Tbl:VWF-Restrict}
\begin{center}
\begin{tabular}{ll}
\hline
 & \hfil Energy\\
\hline
(a) $\sP|\phi_{\text{pair}}\ra$ ($\varphi^{(2)}(\vk)=0$) & $-12.010(8)$\\
(b) $\sP\sL^{S=0}|\phi_{\text{pair}}\ra$ & $-12.459(6)$\\
Exact diagonalization & $-12.5300$\\
\hline
\end{tabular}
\end{center}
\end{table}
%
\par
There are other quantum-number projections to restore symmetries \cite{PIRGMizusaki}. 
The total momentum projection and the lattice symmetry projection restore the translational symmetry and the point group symmetry of lattice, respectively. 
These projections are not adopted in this study. The reason is the following: 
Although the one-body part in eq. (\ref{eq:VWF-one-bodyfull}) can break the translational symmetry (formation of $A$-$B$ sublattice) by the AF symmetry breaking, 
this broken symmetry can be restored by performing the spin projection, 
because a superposition of the rotated wave functions includes a superposition of wave functions translated between $A$ and $B$ sublattice. 
Therefore, after the spin projection, we do not need to introduce other types of symmetry restoration.
\subsection{Calculation of inner product $\la x | \psi\ra$}
For actual VMC calculations, the inner product between a real space configuration $|x\ra$ and a given wave function $|\psi\ra$ is a key quantity. In this section, we explain the way to calculate $\la x|\psi\ra$.
First, we explain the inner product between $|x\ra$ and one-body part $|\phi\ra$ in a general case with $N$ particle ($N$: even). 
The real space configuration $|x\ra$ has the form
\[
  \left\{
\begin{array}{l}
  |x\ra = c_{r_1\gs_1}^{\dag} c_{r_2\gs_2}^{\dag} \cdots c_{r_N\gs_N}^{\dag} |0\ra
\\
  \la x| = \la 0| c_{r_N\gs_N} \cdots c_{r_2\gs_2} \cdots c_{r_1\gs_1}
\end{array}
  \right.
,
\]
and one-body part $|\phi\ra$ has a general pairing functional form
\[
  |\phi\ra = \Biggl[\sum_{i,j=1}^{\Ns} \sum_{\gs,\gs'=\uparrow,\downarrow}
  F_{ij}^{\gs\gs'} c_{i\gs}^{\dag} c_{j\gs'}^{\dag} \Biggr]^{N/2} | 0\ra
.
\]
Here, $F_{ij}^{\gs\gs'}$ is, for example, given by $f_{ij}$ defined in eq. (\ref{eq:VWF-one-bodyfull-real-space}). 
As first pointed out by Bouchaud \etal {} \cite{PfVMC01}, 
the inner product $\la x | \phi\ra$ is given as a Pfaffian of $N\times N$ skew-symmetric matrix. 
We explain this fact below in detail.
\par
Expanding $|\phi\ra$ and picking up nonvanishing terms, we have
\begin{align} \notag
  \la x | \phi\ra 
=
  \la x | &\sum_{\sP} (N/2)! 
\\ \notag
  &\times\prod_{\ell=1}^{N/2} \Bigl(
    F_{r_{\sP(2\ell-1)} r_{\sP(2\ell)}}^{\gs_{\sP(2\ell-1)} \gs_{\sP(2\ell)}}
  - F_{r_{\sP(2\ell)} r_{\sP(2\ell-1)}}^{\gs_{\sP(2\ell)} \gs_{\sP(2\ell-1)}}
  \Bigr)
\\
&\times  \prod_{\ell=1}^{N/2} \Bigl(
    c_{r_{\sP(2\ell-1)}\gs_{\sP(2\ell-1)}}^{\dag} c_{r_{\sP(2\ell)}\gs_{\sP(2\ell)}}^{\dag}
  \Bigr)
  |0\ra
,
\end{align}
where $\sP$ is the permutation of $N$ indices with the condition
\[
  \left\{
\begin{array}{l}
  \sP(2i-1) < \sP(2i)\\[+3pt]
  \sP(1) < \sP(3) < \cdots < \sP(N-1)
\end{array}
  \right.
.
\]
From the power $N/2$ in $|\phi\ra$, the term with the same element of $\prod (F_{rr'}^{\gs\gs'}-F_{r'r}^{\gs'\gs})\prod(c_{r\gs}^\dag c_{r'\gs'}^\dag)$ appears $(N/2)!$ times. 
The commutation relation of fermion operators gives the sign
\[
  \la x |\prod_{\ell=1}^{N/2} \Bigl(
    c_{r_{\sP(2\ell-1)}\gs_{\sP(2\ell-1)}}^{\dag} c_{r_{\sP(2\ell)}\gs_{\sP(2\ell)}}^{\dag}
  \Bigr)
  |0\ra = (-1)^{\sP}
,
\]
where $(-1)^{\sP}$ is the parity of $\sP$. 
Thus, $\la x | \phi\ra$ has the form
\begin{align} \notag
  \la x | \phi\ra &= (N/2)! \sum_{\sP} (-1)^{\sP}
\\ \notag & \qquad \times \prod_{\ell=1}^{N/2} \Bigl(
    F_{r_{\sP(2\ell-1)} r_{\sP(2\ell)}}^{\gs_{\sP(2\ell-1)} \gs_{\sP(2\ell)}}
  - F_{r_{\sP(2\ell)} r_{\sP(2\ell-1)}}^{\gs_{\sP(2\ell)} \gs_{\sP(2\ell-1)}}
  \Bigr)
\\&=
  (N/2)! \, \Pf \mX
,
\end{align}
where $\Pf \mX$ is a Pfaffian of $N\times N$ skew-symmetric matrix $\mX$ with the element
\[
  X_{ij} = F_{r_i r_j}^{\gs_i\gs_j} - F_{r_j r_i}^{\gs_j\gs_i}
.
\]
The linear algebra of skew-symmetric matrix and Pfaffian is described in Appendix \ref{App:PfMat}.
\par
Next, we consider the inner product of a spin projected wave function with the Gutzwiller-Jastrow factor $\sP$: $|\psi\ra=\sP\sL^S|\phi\ra$. 
From eq. (\ref{eq:VWF-SpinProj}), the inner product is given by
\begin{align} \notag
  \la x|\sP\sL^S|\phi\ra = 
  P(x)
  \frac{2S+1}{2}
  \int_{0}^{\pi} \!\!\!\!d\gb
  \, &\sin\beta\, P_{S}(\cos\beta) 
\\
& \times
  \la x|R^y(\gb)|\phi\ra
,
\end{align}
where the condition $S^z=0$ is imposed to $|x\ra$:
\[
  |x\ra = c_{r_1\uparrow}^\dag c_{r_2\uparrow}^\dag \cdots c_{r_{N/2}\uparrow}^\dag
  c_{r_{N/2+1}\downarrow}^\dag c_{r_{N/2+2}\downarrow}^\dag \cdots c_{r_{N}\downarrow}^\dag |0\ra
,
\]
and $|\phi\ra$ has a form
\[
  |\phi\ra = 
    \Biggl[ \sum_{i,j} 
    f_{ij}
    c_{i\uparrow}^\dag c_{j\downarrow}^\dag 
  \Biggr]^{N/2} | 0 \ra
.
\]
The component $\la x|R^y(\gb)|\phi\ra$ can be evaluated by
\begin{align} \notag
  \la x|R^y(\gb)|\phi\ra 
&=
  \la x|
    \Biggl[ \sum_{i,j} 
    f_{ij}
    \Bigl( \cos(\beta/2) c_{i\uparrow}^\dag + \sin(\beta/2) c_{i\downarrow}^\dag \Bigr)
\\ \notag
&\qquad\times
    \Bigl( -\sin(\beta/2) c_{j\uparrow}^\dag + \cos(\beta/2) c_{j\downarrow}^\dag \Bigr)
  \Biggr]^{N/2} \!\!\!\!\!| 0 \ra
\\ \notag
&=
  \la x|
    \Biggl[ \sum_{i,j,\gs,\gs'} 
    F_{ij}^{\gs\gs'} (\beta)
    c_{i\gs}^\dag c_{j\gs'}^\dag 
  \Biggr]^{N/2} | 0 \ra
\\&=
  \Pf \mX(\beta)
,
\end{align}
where the element of skew-symmetric matrix $\mX(\beta)$ is
\[
  X_{ij}(\beta)
=
  \left\{
\begin{array}{@{}l@{\quad}l@{}}
  F_{r_{i}r_{j}}^{\uparrow\uparrow}(\beta) - F_{r_{j}r_{i}}^{\uparrow\uparrow}(\beta)
&
  (i\le N/2 , \, j\le N/2)\\[+2mm]
  F_{r_{i}r_{j}}^{\uparrow\downarrow}(\beta) - F_{r_{j}r_{i}}^{\downarrow\uparrow}(\beta)
&
  (i\le N/2 , \, j> N/2)\\[+2mm]
  F_{r_{i}r_{j}}^{\downarrow\uparrow}(\beta) - F_{r_{j}r_{i}}^{\uparrow\downarrow}(\beta)
&
  (i> N/2 , \, j\le N/2)\\[+2mm]
  F_{r_{i}r_{j}}^{\downarrow\downarrow}(\beta) - F_{r_{j}r_{i}}^{\downarrow\downarrow}(\beta)
&
  (i> N/2 , \, j> N/2)
\end{array}
  \right.
.
\]
Therefore, the inner product $\la x|\sP\sL^S|\phi\ra$ is obtained as
\[
  \la x|\sP\sL^S|\phi\ra = P(x) 
  \frac{2S+1}{2}
  \int_{0}^{\pi} \!\!\!\!d\gb
  \, \sin\beta\, P_{S}(\cos\beta) 
  \Pf \mX(\beta)
.
\]
\section{Optimization Method}
\label{Ch:Opt}
In the procedure of VMC, the wave function optimization is one of the most important tasks. 
In the optimization, we have to keep in mind the following limitations.
\begin{enumerate}[(i)]
\item The estimated value of the cost function (usually the total energy) and its derivatives have the statistical noises by MC samplings.
\item There is a trade off between computational costs and accuracy when one employs the estimation of higher-order derivatives of the energy in the variational parameter space.
\end{enumerate}
In this chapter, first we summarize the basic idea of wave function optimizations by energy minimization. Then, the stochastic reconfiguration (SR) method \cite{SorellaSR}, which Sorella has developed in order to optimize many parameters, is explained in detail.
\subsection{Basic idea of wave function optimization ---Steepest Descent method and Newton method}
We discuss an efficient way of minimizing the energy 
$E_{\vga} = \la \psi_{\vga} | \Ha | \psi_{\vga}\ra / \la \psi_{\vga} | \psi_{\vga}\ra$ estimated from the wave function $| \psi_{\vga}\ra$ with variational parameters $\{\ga_k | k=1,\cdots,p\}$. 
Here $\vga$ denotes the initial 
vector in the $p$-dimensional parameter space.
\par
The energy $E_{\vga+\vgc}$ is expanded up to the second order around $\vga$:
\[
  E_{\vga+\vgc} = E_{\vga} + \sum_{k=1}^{p} g_k \gc_k 
                 + \frac{1}{2} \sum_{k,\ell=1}^{p} h_{k\ell} \gc_k \gc_\ell
                 + \mathcal{O}(\vgc^3),
\]
where $\vgc$ is the vector for parameter variations,
\[
  g_k = \frac{\partial}{\partial \ga_k} E_{\vga}
\quad
  (k=1,\cdots,p)
\]
are the components of the energy gradient vector $\bm{g}$, and
\[
  h_{k\ell} = \frac{\partial^2}{\partial \ga_k \partial \ga_\ell} E_{\vga}
\quad
  (k,\ell=1,\cdots,p)
\]
are the elements of the energy Hessian matrix $\mh$.
\par
With the first order approximation, the steepest decent (SD) method 
gives the updated variational parameter by 
\[
  \ga'_{k} = \ga_{k} + \widebar{\gc}_{k},
\]
where the change from the initial value $\ga_{k}$ should be 
\[ \label{eq:Opt.004}
  \widebar{\gc}_k = -\varDelta t \, g_k
\quad
  (\widebar{\vgc} = -\varDelta t \, \bm{g})
.
\]
Here, $\varDelta t$ is a small constant.
Combination with the second-order information, i.e. the Hessian, leads to the Newton method. By imposing the stationary condition $\partial E_{\vga} / \partial \ga_k = 0$ ($k=1,\cdots,p$), 
the best parameter change $\widebar{\vgc}$ is obtained by
\[ \label{eq:Opt.005}
  \widebar{\gc}_k = -\sum_{\ell=1}^{p} h_{k\ell}^{-1} g_\ell
\quad
  (\widebar{\vgc} = - \mh^{-1} \bm{g}).
\]
\par
Let us generalize eqs. (\ref{eq:Opt.004}) and (\ref{eq:Opt.005}) with suitably chosen nonsingular matrix $\mX$:
\[ \label{eq:Opt.006}
  \widebar{\gc}_k = -\sum_{\ell=1}^{p} X_{k\ell}^{-1} g_\ell
\quad
  (\widebar{\vgc} = - \mX^{-1} \bm{g}).
\]
Equations (\ref{eq:Opt.004}) and (\ref{eq:Opt.005}) are reduced from eq. (\ref{eq:Opt.006}) by setting $X_{k\ell}=\delta_{k\ell} / \varDelta t$ and $X_{k\ell} = h_{k\ell}$, respectively.
\par
As long as the energy gradient $\bm{g}$ is estimated with the mathematically correct formula, the parameter change $\widebar{\vgc}$ will converge at the minimum or at a stationary point irrespective of the choice of $\mX$. 
However, the computational efficiency strongly depends on the choice
of $\mX$ and $\mX$ should be chosen to accelerate the optimization 
within computational stability.
\subsection{Stochastic Reconfiguration method}
Sorella has developed the SR method \cite{SorellaSR}, which offers a simple but very stable optimization method. 
In order to deal with a large number of variational parameters and optimize all the parameters simultaneously, we employ the SR method in this studies.
First, we introduce the normalized wave function
\[
  |\widebar{\psi}_{\vga}\ra = \frac{1}{\sqrt{ \la \psi_{\vga}|\psi_{\vga}\ra }} |\psi_{\vga}\ra.
\]
Then the expansion of $|\widebar{\psi}_{\vga+\vgc}\ra$ up to the first order around $\vga$ is
\[
  |\widebar{\psi}_{\vga+\vgc}\ra = |\widebar{\psi}_{\vga}\ra 
  + \sum_{k=1}^{p} \gc_k |\widebar{\psi}_{k\vga}\ra
  + \mathcal{O}(\vgc^2),
\]
where $|\widebar{\psi}_{k\vga}\ra$ ($k=1,\cdots,p$) are the derivatives of $|\widebar{\psi}_{\vga}\ra$:
\begin{align}\notag
  &|\widebar{\psi}_{k\vga}\ra = \frac{\partial}{\partial \ga_k} |\widebar{\psi}_{\vga}\ra
\\ \label{eq:Opt-SR-049}
&=
  \frac{1}{\sqrt{\la \psi_{\vga}|\psi_{\vga}\ra }}
  \biggl(
    \frac{\partial}{\partial \ga_k} |\psi_{\vga}\ra
  - \frac{\la \psi_{\vga}|(\partial / \partial \ga_k)|\psi_{\vga}\ra}{\la \psi_{\vga}|\psi_{\vga}\ra} 
    |\psi_{\vga}\ra
  \biggr).
\end{align}
The wave function set $\{ |\widebar{\psi}_{k\vga}\ra | k=1,\cdots,p \}$ forms nonorthogonal basis in the $p$-dimensional parameter space. 
The norm of the variation between $|\widebar{\psi}_{\vga}\ra$ and $|\widebar{\psi}_{\vga+\vgc}\ra$ is
\begin{align}\notag
  \varDelta_{\text{norm}}^{2} 
&=
  \Bigl\Vert |\widebar{\psi}_{\vga+\vgc}\ra - |\widebar{\psi}_{\vga}\ra \Bigr\Vert ^2
\\ \label{eq:Opt.010}
&=
  \sum_{k,\ell=1}^{p} \gc_k\gc_\ell \la\widebar{\psi}_{k\vga}|\widebar{\psi}_{\ell\vga}\ra
=
  \sum_{k,\ell=1}^{p} \gc_k\gc_\ell S_{k\ell}.
\end{align}
Since $S_{k\ell} = \la\widebar{\psi}_{k\vga}|\widebar{\psi}_{\ell\vga}\ra$ is the overlap matrix in the parameter space, 
$\mS$ becomes positive definite even with a finite number of samples. 
Equation (\ref{eq:Opt.010}) shows that $\mS$ is the metric matrix in the parameter space.
\par
The SR method chooses $\mS$ as the matrix $\mX$ in eq. (\ref{eq:Opt.006}), namely
\[ \label{eq:Opt.011}
  \widebar{\gc}_k = -\varDelta t \sum_{\ell=1}^{p} S_{k\ell}^{-1} g_\ell
\quad
  (\widebar{\vgc} = -\varDelta t \, \mS^{-1}\bm{g})
,
\]
where $\varDelta t$ is a small constant.
As the overlap matrix $\mS$ does not have any information about the energy Hessian, the SR method is close to the SD method. 
The main difference is that the SR method takes into account the variation of the wave function. We can derive eq. (\ref{eq:Opt.011}) by minimizing the functional 
$
  \mathcal{F}_{\text{SR}}
= \varDelta E_{\text{lin.}} + \lambda \varDelta_{\text{norm}}^{2}
$
with a Lagrange multiplier $\lambda$. Here $\varDelta E_{\text{lin.}} = \sum_{k} g_k\gc_k$ is the linear change of the energy. The stationary condition $\partial \mathcal{F}_{\text{SR}} / \partial \gc_k=0$ ($k=1,\cdots,p$) leads to the SR formula (\ref{eq:Opt.011}) with $\varDelta t = (2\lambda)^{-1}$. 
The SD method can be obtained in a similar way. We can derive eq. (\ref{eq:Opt.004}) with $\varDelta t = (2\lambda)^{-1}$ by minimizing the functional 
$
  \mathcal{F}_{\text{SD}}
= \varDelta E_{\text{lin.}} + \lambda \varDelta_{\text{SD}}^{2}
$
with 
$
  \varDelta_{\text{SD}}^{2} = 
  \sum_{k} \gc_{k}^2
,
$
where $\varDelta_{\text{SD}}$ is the Cartesian distance in the parameter space.
The advantage of the SR method compared with the SD method is the following: Sometimes small change of the variational parameters corresponds to a large change of the wave function, and conversely a large change of the variational parameters corresponds to a small change of the wave function. 
This leads to uncontrolled changes of the wave function if one takes $S_{k\ell}=\delta_{k\ell}$ (SD method). 
When the change of the wave function exceeds a threshold, the iteration  for the optimization becomes unstable. To suppress this instability, one needs to keep $\varDelta t$ small enough for the event of the largest change of the wave function. 
If one can control change of the wave function, $\varDelta t$ can be taken large. 
The SR method takes into account this effect through a better definition of the distance $\varDelta_{\text{norm}}$. 
Thus, the SR method is more stable than the SD method. 
Finally, we can choose larger $\varDelta t$ to accelerate the convergence.
\subsection{Stabilization of SR method}
In the VMC calculation, it is important to optimize the wave function stably with a small number of samples. 
The main instability in the SR method comes from the overlap matrix $\mS$. 
Though the overlap matrix $\mS$ is positive definite even with a finite number of samples, the inverse matrix $\mS^{-1}$ amplifies the statistical noise in the energy gradient $\bm{g}$ when the ratio of the maximum eigenvalue and the minimum eigenvalue becomes extremely large.
The statistical noise of MC sampling and a variety in dependence on each parameters enlarge this ratio.
\par
To stabilize the SR method, we apply two techniques \cite{CasulaJCP, SorellaJCP}:
\begin{enumerate}[(i)]
\item Modification of diagonal elements in $\mS$,
\item Truncation of redundant directions in the parameter space.
\end{enumerate}
\subsubsection{Modification of diagonal elements in $\mS$}
As we have already mentioned, deformation of $\mX$ in eq. (\ref{eq:Opt.006}) does not change the optimal parameters. Then we follow the stabilization method in ref. \citen{SorellaJCP} by modifying diagonal elements in $\mS$:
\[ \label{eq:Opt-SRMod-01}
  S_{kk} \to (1+\varepsilon) S_{kk},
\]
where $\varepsilon\ll 1$ is a small constant. 
This modification preserves the positive definite property, because the sum of two positive definite matrices $S_{k\ell}$ and $\varepsilon \delta_{k\ell} S_{k\ell}$ remains a positive definite matrix. 
This modified matrix pulls up extremely small eigenvalues and suppresses fluctuations in the SR iteration.
\par
As in the stabilization of the ordinary Newton method, it is possible to add a uniform constant $\varepsilon$ to diagonal elements ($S_{kk}\to S_{kk}+\varepsilon$). This also stabilizes the SR method, but the convergence becomes slower than the former, because this modification does not take account of the metric in the parameter space. 
Though this stabilization in eq. (\ref{eq:Opt-SRMod-01}) suppresses fluctuations in a large part of the variational parameter space, it becomes inefficient in some cases. 
When the ratio of the maximum value and the minimum value of the diagonal elements becomes extremely large, the ill part is not modified efficiently by the additional matrix $\varepsilon \delta_{k\ell} S_{k\ell}$. 
In order to stabilize the SR method with large statistical noises further, we combine the modification in eq. (\ref{eq:Opt-SRMod-01}) with a truncation technique as discussed in the next part.
\subsubsection{Truncation of redundant directions}
Casula \etal {} have introduced a truncation technique for irrelevant variational parameters to stabilize the SR method \cite{CasulaJCP}. They have directly truncated some parameters. Here, we introduce a better truncation procedure by considering the eigenvalues of the overlap matrix $\mS$.
\par
As $\mS$ is a $p$-dimensional positive definite symmetric matrix, we can diagonalize $\mS$ by an orthogonal matrix $\mU$:
\[ \label{eq:Opt.014}
  \sum_{k,\ell=1}^{p} U_{ki}U_{\ell j} S_{k\ell} = \lambda_i \delta_{ij}
\Longleftrightarrow 
\left\{
\begin{array}{@{}c@{}}\displaystyle
  S_{k\ell} = \sum_{i=1}^{p} \lambda_{i} U_{ki} U_{\ell i}
\\[+4mm]
\displaystyle
  S_{k\ell}^{-1} = \sum_{i=1}^{p} \frac{1}{\lambda_i} U_{ki}U_{\ell i}
\end{array}
\right.,
\]
where $\lambda_{i} > 0$ ($i=1,\cdots,p$) are the eigenvalues and arranged in descending order ($\lambda_{1}\ge\lambda_{2}\ge\cdots\ge\lambda_{p}$). 
Then we apply an orthogonal transformation to the parameter change $\vgc$:
\[ \label{eq:Opt.015}
  x_k = \sum_{k=1}^{p} \gc_k U_{ki}
\ \Longleftrightarrow \ 
  \gc_k = \sum_{i=1}^{p} U_{ki} x_i.
\]
From eqs. (\ref{eq:Opt.010}), (\ref{eq:Opt.014}), and (\ref{eq:Opt.015}), we obtain the following:
\[
  \varDelta_{\text{norm}}^{2} = \sum_{k,\ell,i=1}^{p} \gc_k \gc_\ell \lambda_i U_{ki}U_{\ell i}
= \sum_{i=1}^{p} \lambda_i x_i^2.
\]
This means that the variation in the direction $x_i$ where $\lambda_i/\lambda_1 < \varepsilon_{\text{wf}}$ is satisfied is redundant in the range of relative accuracy $\varepsilon_{\text{wf}}$. 
Moreover, this direction brings an instability ($\sim 1/\varepsilon_{\text{wf}}$) to $\mS^{-1}$.
Therefore, we can control the relative accuracy of $\varDelta_{\text{norm}}^{2}$ and stability of $\mS^{-1}$ by tuning $\varepsilon_{\text{wf}}$. 
We truncate the direction $\{x_i|i=q+1,\cdots,p\}$ which satisfy $\lambda_i/\lambda_1 < \varepsilon_{\text{wf}}$. The SR formula (\ref{eq:Opt.011}) is changed into 
\begin{gather}
  \widebar{\gc}_k = - \varDelta t \sum_{\ell=1}^{p} S_{k\ell}^{-1} g_\ell = -\varDelta t \sum_{k=1}^{p} \sum_{i=1}^{p} \frac{1}{\lambda_{i}}U_{ki}U_{\ell i}g_\ell
\\[-2mm]
\Big\Downarrow \ (\text{Truncation}) \notag
\\[-2mm]
  \widebar{\gc}_k = -\varDelta t \sum_{\ell=1}^{p} \sum_{i=1}^{q} \frac{1}{\lambda_{i}}U_{ki}U_{\ell i}g_\ell.
\end{gather}
\par
By introducing the truncation parameter $\varepsilon_{\text{wf}}$, we can control both the accuracy and the stability of the SR optimization. 
The truncation by decomposing $\mS$ into the orthogonal matrix $\mU$ (eq. (\ref{eq:Opt.014})) is essentially equivalent to the singular value decomposition (SVD) method \cite{NumRec} to handle ill matrices. 
In the optimization method of variational wave functions, the SVD has been adopted in the linear method developed by Nightingale and Melik-Alaverdian, which uses some information of the Hessian \cite{VMCOptLM02, VMCOptLM01}.
\subsubsection{Practical parameters in SR method}
In order to perform the stable SR optimization with small MC samples ($4\times 10^3$-$10^4$ samples), we practically choose the parameters in the SR method as $\varDelta t = 0.1$, $\varepsilon = 0.2$, and $\varepsilon_{\text{wf}} = 0.001$. The optimized variational parameters are typically %
obtained by averaging the parameters over $100$ optimization steps after $500$-$5000$ steps.
\subsection{Derivative operator}
In order to perform the SR optimization, we need to evaluate the overlap matrix $\mS$ and the energy gradient $\bm{g}$. 
We follow a standard way in continuum systems by introducing derivative operators \cite{VMCOptNW01,VMCOptNW02}. In lattice models, Sorella have derived representations of these operators \cite{SorellaSR, SorellaSRH}.
\par
The wave function considered here is a quantum-number projected one with the Gutzwiller-Jastrow factor 
\[
  |\psi_{\vga}\ra = \sP_{\vga} \sL |\phi_{\vga}\ra
=
  \sP_{\vga}\sum_{n}w_n |\phi_{\vga}^{(n)}\ra
,
\]
where the summation $\sum_{n}$ and weights $w_n$ come from the quantum-number projection $\sL$. 
Here, the Gutzwiller-Jastrow factor 
$\sP_{\vga}$ and 
the transformed one-body parts 
$|\phi_{\vga}^{(n)}\ra$ have variational parameters. We assume that the parameter dependence in $\sP_{\vga}$ and $|\phi_{\vga}^{(n)}\ra$ is separated.
\par
First, we introduce operators $\sO_{k}$ ($k=1,\cdots,p$) which are diagonal in real space configurations $|x\ra$. $\sO_{k}$ is defined as
\[
  \sO_k = \sum_{x} |x\ra \biggl[
    \frac{1}{\la x|\psi_{\vga}\ra} \frac{\partial}{\partial \ga_k} \la x|\psi_{\vga}\ra
  \biggr] \la x|
=
  \sum_{x} |x\ra O_k(x) \la x|
\]
and satisfies the following relations:
\begin{gather}
  \la x|\sO_k|\psi_{\vga}\ra = \frac{\partial}{\partial\ga_k} \la x|\psi_{\vga}\ra
,
\\ 
  \sO_k|\psi_{\vga}\ra = \sum_{x} |x\ra \frac{\partial}{\partial\ga_k} \la x|\psi_{\vga}\ra
,
\\
  \la \sO_k\ra = \frac{\la \psi_{\vga}|\sO_k|\psi_{\vga}\ra}{\la \psi_{\vga}|\psi_{\vga}\ra}
=
  \frac{\la \psi_{\vga}| (\partial / \partial \ga_k) |\psi_{\vga}\ra}{\la \psi_{\vga}|\psi_{\vga}\ra}
.
\end{gather}
The derivative of the normalized wave function (eq. (\ref{eq:Opt-SR-049})) is rewritten as
\[
  |\widebar{\psi}_{k\vga}\ra 
=
  \frac{1}{\sqrt{\la \psi_{\vga}|\psi_{\vga}\ra }}
  ( \sO_k - \la \sO_k \ra )
    |\psi_{\vga}\ra
,
\]
and the overlap matrix $\mS$ is obtained as
\[
  S_{k\ell} 
=
  \la \sO_k\sO_\ell\ra - \la\sO_k\ra\la\sO_\ell\ra
.
\]
The energy gradient $\bm{g}$ is also written as
\begin{align} \notag
  g_k &= \frac{\partial}{\partial \ga_k} \la \widebar{\psi}_{\vga}| \Ha |\widebar{\psi}_{\vga}\ra
=
  \la \widebar{\psi}_{k\vga}| \Ha |\widebar{\psi}_{\vga}\ra
+ \la \widebar{\psi}_{\vga}| \Ha |\widebar{\psi}_{k\vga}\ra
\\
&=
  2 \la\Ha\sO_k\ra - 2\la\Ha\ra\la\sO_k\ra
.
\end{align}
\par
Next, we derive expressions for $O_{k}(x)$. The Gutzwiller-Jastrow factor has an exponential form
\[
  \sP_{\vga} = \exp \biggl[ - \sum_{k} \ga_k \varTheta_k \biggr]
,
\]
where $\ga_k$ are variational parameters and $\varTheta_k$ are diagonal operators for real space configurations ($\varTheta_k |x\ra = \varTheta_k(x)|x\ra$). 
$O_k(x)$ corresponding to $\sP_{\vga}$ is obtained as
\[
  O_k(x) = -\varTheta_k(x)
.
\]
Since the inner product $\la x|\phi_{\vga}^{(n)}\ra$ is evaluated by the Pfaffian $\Pf\mX_{\vga}^{(n)}$, $O_k(x)$ corresponding to the one-body part is related to the derivative of the Pfaffian. The result is given as
\begin{align} \notag
  O_k(x)
&=
  \frac{1}{\la x|\psi_{\vga}\ra} \frac{\partial}{\partial \ga_k} \la x|\psi_{\vga}\ra
=
  \frac{1}{\la x|\sL|\phi_{\vga}\ra} \frac{\partial}{\partial \ga_k} \la x|\sL|\phi_{\vga}\ra
\\ \notag
&=
  \frac{1}{\la x|\sL|\phi_{\vga}\ra} \frac{\partial}{\partial \ga_k} \sum_n w_n \la x|\phi_{\vga}^{(n)}\ra
\\ \notag
&=
  \frac{1}{\la x|\sL|\phi_{\vga}\ra} \sum_n w_n \frac{\partial}{\partial \ga_k} \Pf\mX_{\vga}^{(n)}
\\
&=
  \frac{\displaystyle \sum_n w_n \frac{1}{2} \Pf\mX_{\vga}^{(n)} 
  \Tr \biggl[ \mX_{\vga}^{(n)-1} \frac{\partial}{\partial\ga_k} \mX_{\vga}^{(n)}\biggr]
       }{\displaystyle \sum_n w_n \Pf\mX_{\vga}^{(n)}}
.
\end{align}
The overlap matrix $\mS$ and the energy gradient $\bm{g}$ are evaluated under the VMC sampling by using the above formulae. 
For example, if $|\phi_{\vga}\ra$ is given by $|\phi_{\text{pair}}\ra$ defined by eq. (\ref{eq:VWF-one-bodyfull}), 
$\partial \mX_{\vga}^{(n)} / \partial\ga_k$ contains derivatives $\partial f_{ij}/ \partial \varphi^{(1)}(\vk)$ and $\partial f_{ij}/ \partial \varphi^{(2)}(\vk)$ through eq. (\ref{eq:VWF-one-bodyfull-real-space}).
\section{Results}
\label{Ch:Bench}
In this section, we apply our improved wave functions to variational calculations for the two-dimensional Hubbard model. 
We compare variational results obtained from our wave functions with 
unbiased results obtained by the ED method and the AFQMC method \cite{AFQMC}.
\subsection{Comparison with exact diagnalization and energy variance}
%
\begin{table*}[t]
\caption{Comparison of variational energies 
obtained from different wave functions and the exact energy calculated by ED 
for $L=4$, $t'/t=0$, and $n=1$. The numbers in parentheses are 
the statistical errors in the last digits.}
\label{Tbl:Res-Rerror}
\begin{center}
\begin{tabular}{r@{\ \,}l@{\;}l@{\;}l@{\;}l@{\;}l}
\hline
 & \multicolumn{1}{c}{$U/t=4$} & \multicolumn{1}{c}{$5$} & \multicolumn{1}{c}{$10$} & \multicolumn{1}{c}{$20$} & \multicolumn{1}{c}{$35$} \\
\hline
$| \phi_{\text{AF}} \rangle$ & 
$-12.925(0)$ & 
$-10.979(0)$ & 
$-6.1089(0)$ & 
$-3.1610(0)$ & 
$-1.8212(0)$
\\[+1mm]
$\mathcal{P}_{\text{G}} | \phi_{\text{AF}} \rangle$ & 
$-14.09(1)$ & 
$-11.82(2)$ & 
$-6.41(1)$ & 
$-3.24(1)$ & 
$-1.84(2)$
\\[+1mm]
$\mathcal{P}_{\text{d-h}}\mathcal{P}_{\text{G}} | \phi_{\text{AF}} \rangle$ & 
$-14.204(1)$ & 
$-11.951(1)$ & 
$-6.551(1)$ & 
$-3.42(1)$ & 
$-1.978(5)$
\\[+1mm]
$\mathcal{L}^{S=0} | \phi_{\text{AF}} \rangle$ & 
$-13.735(5)$ & 
$-11.775(5)$ & 
$-6.72(1)$ & 
$-3.54(1)$ & 
$-2.04(2)$
\\[+1mm]
$\mathcal{P}_{\text{d-h}}\mathcal{P}_{\text{G}} \mathcal{L}^{S=0}| \phi_{\text{AF}} \rangle$ & 
$-14.437(2)$ & 
$-12.356(1)$ & 
$-6.946(1)$ & 
$-3.63(1)$ & 
$-2.10(1)$
\\[+1mm]
$\mathcal{P}_{\text{J}}\mathcal{P}_{\text{d-h}}^{\text{ex.}} \mathcal{P}_{\text{G}} \mathcal{L}^{S=0}| \phi_{\text{AF}} \rangle$ & 
$-14.443(2)$ & 
$-12.363(2)$ & 
$-6.952(4)$ & 
$-3.629(6)$ & 
$-2.101(4)$
\\[+1mm]
$\mathcal{P}_{\text{J}}\mathcal{P}_{\text{d-h}}^{\text{ex.}} \mathcal{P}_{\text{G}}| \phi_{\text{pair}} \rangle$ & 
$-14.278(1)$ & 
$-12.10(1)$ & 
$-6.738(7)$ & 
$-3.486(7)$ & 
$-2.009(8)$
\\[+1mm]
$\mathcal{P}_{\text{J}}\mathcal{P}_{\text{d-h}}^{\text{ex.}} \mathcal{P}_{\text{G}} \mathcal{L}^{S=0}| \phi_{\text{pair}} \rangle$ & 
$-14.512(3)$ & 
$-12.459(6)$ & 
$-7.050(2)$ & 
$-3.693(2)$ & 
$-2.135(4)$
\\[+1mm]
Exact (ED) & 
$-14.5935$ & 
$-12.5300$ & 
$-7.13239$ & 
$-3.76124$ & 
$-2.18092$
\\
\hline
\end{tabular}
\end{center}
%
\caption{%
Energy $E$, double occupancy $D$, and spin structure factor $S(\pi,\pi)$ of ED and VMC results. The variational wave function is $\mathcal{P}_{\text{J}}\mathcal{P}_{\text{d-h}}^{\text{ex.}} \mathcal{P}_{\text{G}} \mathcal{L}^{S=0} | \phi_{\text{pair}} \rangle$. The numbers 
in parentheses are the statistical errors in the last digits.}
\label{Tbl:Res-tperror}
\begin{center}
{\scriptsize
\begin{tabular}{l@{}l@{\,}l@{\,}l@{\,}l@{\,}l@{\,}l@{\,}l@{\,}l@{\,}l}
\hline
 & $t'/t=0$ &&& $t'/t=-0.3$ &&& $t'/t=-0.5$ && \\\hline
 \hfil$U/t$&\hfil$4$&\hfil$8$&\hfil$10$&\hfil$4$&\hfil$8$&\hfil$10$&\hfil$4$&\hfil$8$&\hfil$10$\\
\hline
$E_{\text{ED}}$           &$-14.5935$&$-8.63871$&$-7.13239$&$-14.6322$&$-8.66093$&$-7.15550$&$-14.7305$&$-8.71446$&$-7.21193$\\
$E_{\text{VMC}}$          &$-14.512(3)$&$-8.569(2)$&$-7.050(2)$&$-14.531(2)$&$-8.571(2)$&$-7.048(3)$&$-14.561(3)$&$-8.578(2)$&$-7.017(6)$ \\
$D_{\text{ED}}$           &$\phantom{-}0.14389$&$\phantom{-}0.05664$&$\phantom{-}0.03911$&$\phantom{-}0.14410$&$\phantom{-}0.05661$&$\phantom{-}0.03910$&$\phantom{-}0.14463$&$\phantom{-}0.05650$&$\phantom{-}0.03906$\\
$D_{\text{VMC}}$          &$\phantom{-}0.1441(3)$&$\phantom{-}0.0570(1)$&$\phantom{-}0.0386(2)$&$\phantom{-}0.1442(2)$&$\phantom{-}0.0569(3)$&$\phantom{-}0.0394(4)$&$\phantom{-}0.1448(3)$&$\phantom{-}0.0573(2)$&$\phantom{-}0.0413(2)$\\
$S_{\text{ED}}(\pi,\pi)$  &$\phantom{-}0.63517$&$\phantom{-}1.21722$&$\phantom{-}1.31045$&$\phantom{-}0.61656$&$\phantom{-}1.20136$&$\phantom{-}1.28914$&$\phantom{-}0.56647$&$\phantom{-}1.15483$&$\phantom{-}1.22434$\\
$S_{\text{VMC}}(\pi,\pi)$ &$\phantom{-}0.674(3)$&$\phantom{-}1.250(2)$&$\phantom{-}1.368(2)$&$\phantom{-}0.664(1)$&$\phantom{-}1.260(2)$&$\phantom{-}1.367(3)$&$\phantom{-}0.648(1)$&$\phantom{-}1.261(2)$&$\phantom{-}1.371(4)$ \\
\hline
\end{tabular}
}
\end{center}
\end{table*}
First, we compare the energy of different variational wave functions with the ED results. 
Table \ref{Tbl:Res-Rerror} shows the results for 
the system $L=4$, $t'/t=0$, $U/t=4$, and $n=1$. 
The one-body part denoted by $|\phi_{\text{AF}}\ra$ is the mean-field state which 
diagonalizes the Hamiltonian (\ref{eq:VWF-HamMF}) 
with $\varDelta_{\text{SC}}^{a}(\vk) = \varDelta_{\text{SC}}^{b}(\vk) = 0$
and with $\varDelta_{\text{AF}}$ being optimized as a variational parameter of $|\phi_{\text{AF}}\ra$. 
\par
\begin{figure}[t]
\vspace*{-5mm}
\centering
\includegraphics[width=0.45\textwidth]{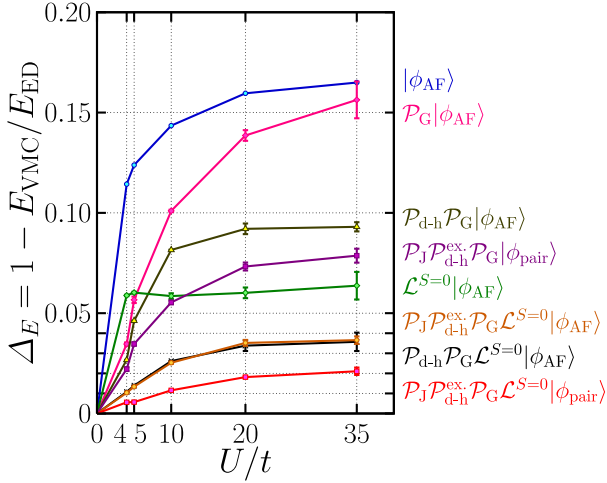}
\caption{%
(Color online) Relative accuracy of different wave functions for $L=4$, $t'/t=0$, and $n=1$. The accuracy $\varDelta_{E}=1-E_{\text{VMC}}/E_{\text{ED}}$ denotes relative difference of the variational energy $E_{\text{VMC}}$ and the exact value $E_{\text{ED}}$. 
}
\label{fig:Res-Rerror}
\end{figure}
\par
Figure \ref{fig:Res-Rerror} shows the relative accuracy of the above results. The spin-projected generalized pairing wave function with the Gutzwiller-Jastrow factor has the best accuracy. The spin projection acts efficiently in each wave function. Restoration of spin rotational symmetry and filtering out of excited states with other spin quantum-numbers are crucial to improve variational wave functions. 
The Jastrow factor and the extension of the doublon-holon correlation factor do not seem to lower the energy substantially.
Table \ref{Tbl:Res-tperror} shows the comparison at nonzero $t'/t$ 
for the same lattice size. 
The accuracy slightly declines as the frustration increases. The double occupancy $D$ in Table \ref{Tbl:Res-tperror} is defined as 
$
  D = 
     (1/\Ns) \sum_{i} \la n_{i\uparrow} n_{i\downarrow} \ra
$.
\par
Table \ref{Tbl:Res-Var} shows the comparison of the energy and the energy variance $(\la \Ha^2\ra-\la\Ha\ra^2)/\la\Ha\ra^2$. 
It shows that the improvement is brought step by step by implementing each refined treatment of $|\phi_{\text{pair}}\ra$, $\sL^{S=0}$, $\mathcal{P}_{\text{G}}$, $\mathcal{P}_{\text{d-h}}^{\text{ex.}}$, and $\mathcal{P}_{\text{J}}$.
%
\begin{table}[t]
\vspace*{-7.5mm}
\caption{
Comparison of energies and energy variances 
obtained from wave functions with different components and the exact energy calculated by ED 
for $L=4$, $t'/t=0$, $U/t=4$, and $n=1$. The numbers in parentheses are the statistical errors in the last digits.}
\label{Tbl:Res-Var}
\begin{center}
\begin{tabular}{rlc}
\hline
 & \hfil$E$ & $(\la \Ha^2\ra-\la\Ha\ra^2)/\la\Ha\ra^2$\\
\hline
$|\phi_{\text{pair}}\ra$ &
$-13.006(6)$ & $0.0350(2)$\\
$\mathcal{L}^{S=0}|\phi_{\text{pair}}\ra$ &
$-13.831(4)$ & $0.0231(6)$\\
$\mathcal{P}_{\text{J}}\mathcal{P}_{\text{d-h}}^{\text{ex.}}\mathcal{P}_{\text{G}}|\phi_{\text{pair}}\ra$ &
$-14.278(1)$ & $0.0096(3)$\\
$\mathcal{P}_{\text{G}}\mathcal{L}^{S=0}|\phi_{\text{pair}}\ra$ &
$-14.489(2)$ & $0.0032(2)$\\
$\mathcal{P}_{\text{d-h}}^{\text{ex.}}\mathcal{P}_{\text{G}}\mathcal{L}^{S=0}|\phi_{\text{pair}}\ra$&
$-14.509(2)$ & $0.0028(1)$\\
$\mathcal{P}_{\text{J}}\mathcal{P}_{\text{d-h}}^{\text{ex.}}\mathcal{P}_{\text{G}}\mathcal{L}^{S=0}|\phi_{\text{pair}}\ra$ &
$-14.512(3)$ & $0.0028(3)$\\
Exact (ED) &
$-14.5935$ & $0$\\
\hline
\end{tabular}
\end{center}
\end{table}
%
\subsection{Size dependence}
%
\begin{figure}[t]
\centering
\includegraphics[width=0.45\textwidth]{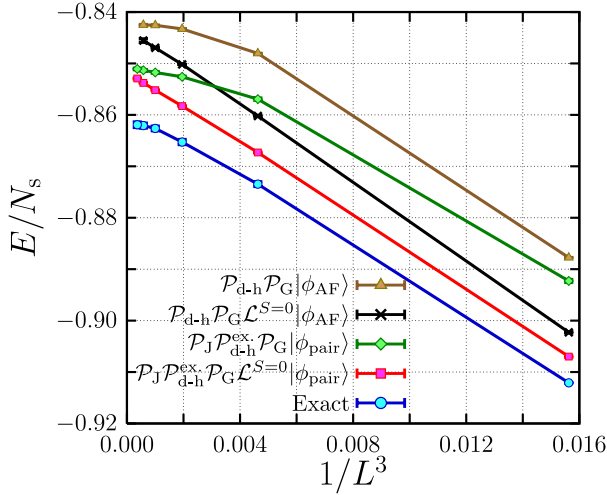}
\caption{%
(Color online) Total energy $E/\Ns$ as a function of $1/L^3$ for 
$t'/t=0$, $U/t=4$, $n=1$. The exact values are calculated by ED ($L=4$) and AFQMC ($L=6,8,10,12,14$).
Error bars are comparable to the symbol size.
}
\label{fig:Res-SizeEn}
\end{figure}
\par
Next, we examine size dependence by comparing with AFQMC results, which are exact within the statistical accuracy. Figure \ref{fig:Res-SizeEn} shows the size dependence of total energy 
per site $E/\Ns$, whose difference from the thermodynamic limit $L \to \infty$
scales proportional to $1/L^3$ 
as is derived from spin-wave theory \cite{Huse}. The spin-projected generalized pairing wave 
function with the Gutzwiller-Jastrow factor has the best variational energy. The energy gain by the spin projection becomes smaller when the system size increases, because the spin 
projection lowers the energy by filtering out higher-energy excited states with other spin quantum-numbers while the energy is not efficiently lowered when the energy of excitations 
belonging to the same quantum number as the ground state 
becomes close to the ground-state energy in larger sizes. 
In finite size calculations, however, the spin projection is still useful to obtain 
better wave functions on a quantitative level. 
Furthermore, 
the energy of the spin-projected wave functions is better 
scaled by $1/L^3$ than that of unprojected cases, which makes the 
extrapolation to the thermodynamic limit easier. 
\par
Introducing a large number of variational parameters into the one-body part efficiently improves the wave function and allows us to go beyond the simple AF order generated from the mean-field treatment. 
We remark that the additional parameters for $|\phi_{\text{AF}}\ra$ ($\varDelta_{\text{SC}}^{a(b)}(\vk)=\varDelta_{\text{SC}}^{a(b)} (\cos k_x-\cos k_y)$ and $\mu$ in eq. (\ref{eq:VWF-HamMF})) do not improve the wave function in this parameter region. We confirm this fact in the small system $L=4$ and $L=6$.
\begin{figure}[t]
\centering
\includegraphics[width=0.45\textwidth]{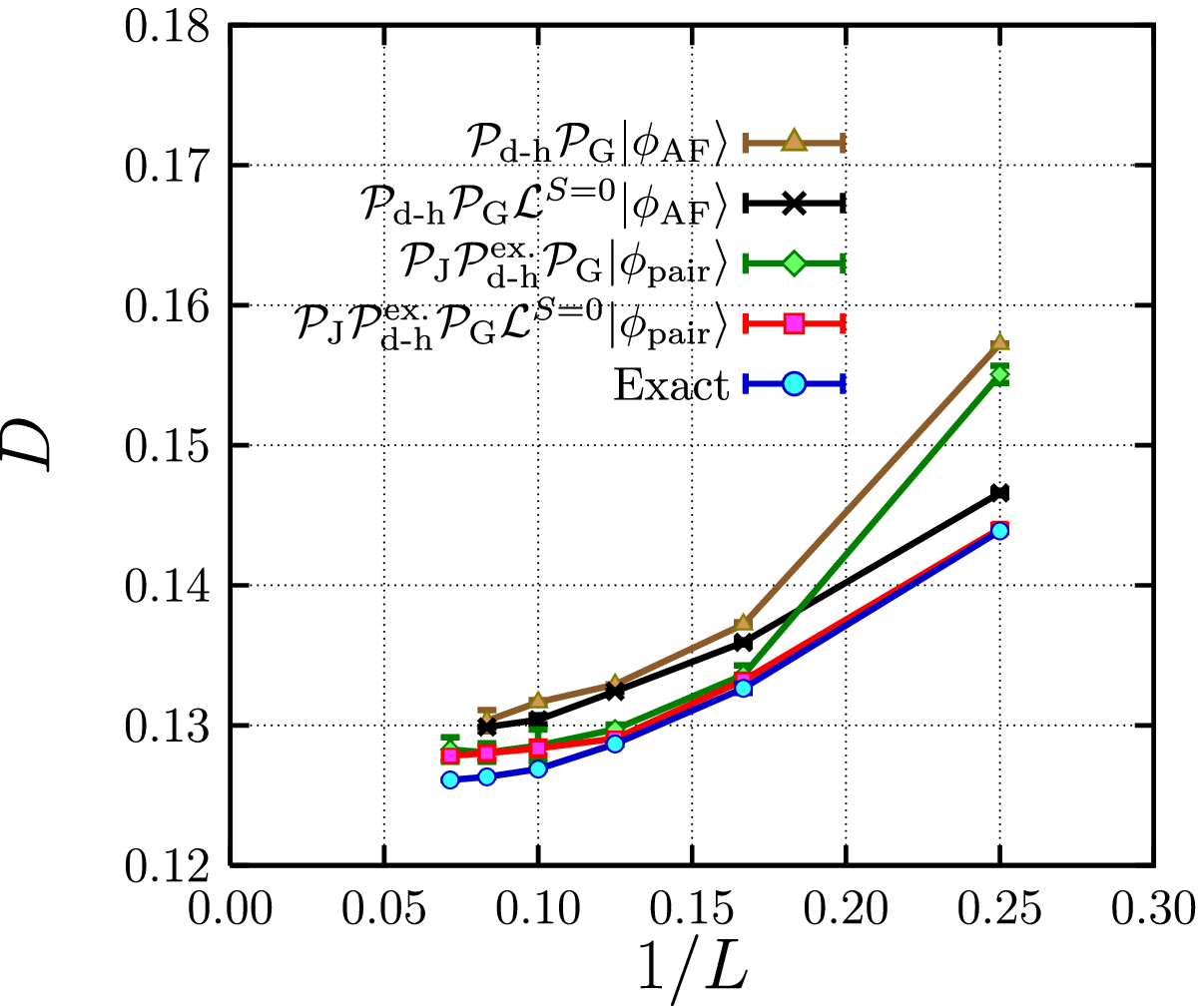}
\caption{%
(Color online) Size dependence of double occupancy. The exact values are calculated by the same way as Fig. \ref{fig:Res-SizeEn}.
}
\label{fig:Res-SizeDo}
\vspace*{5mm}
\centering
\includegraphics[width=0.45\textwidth]{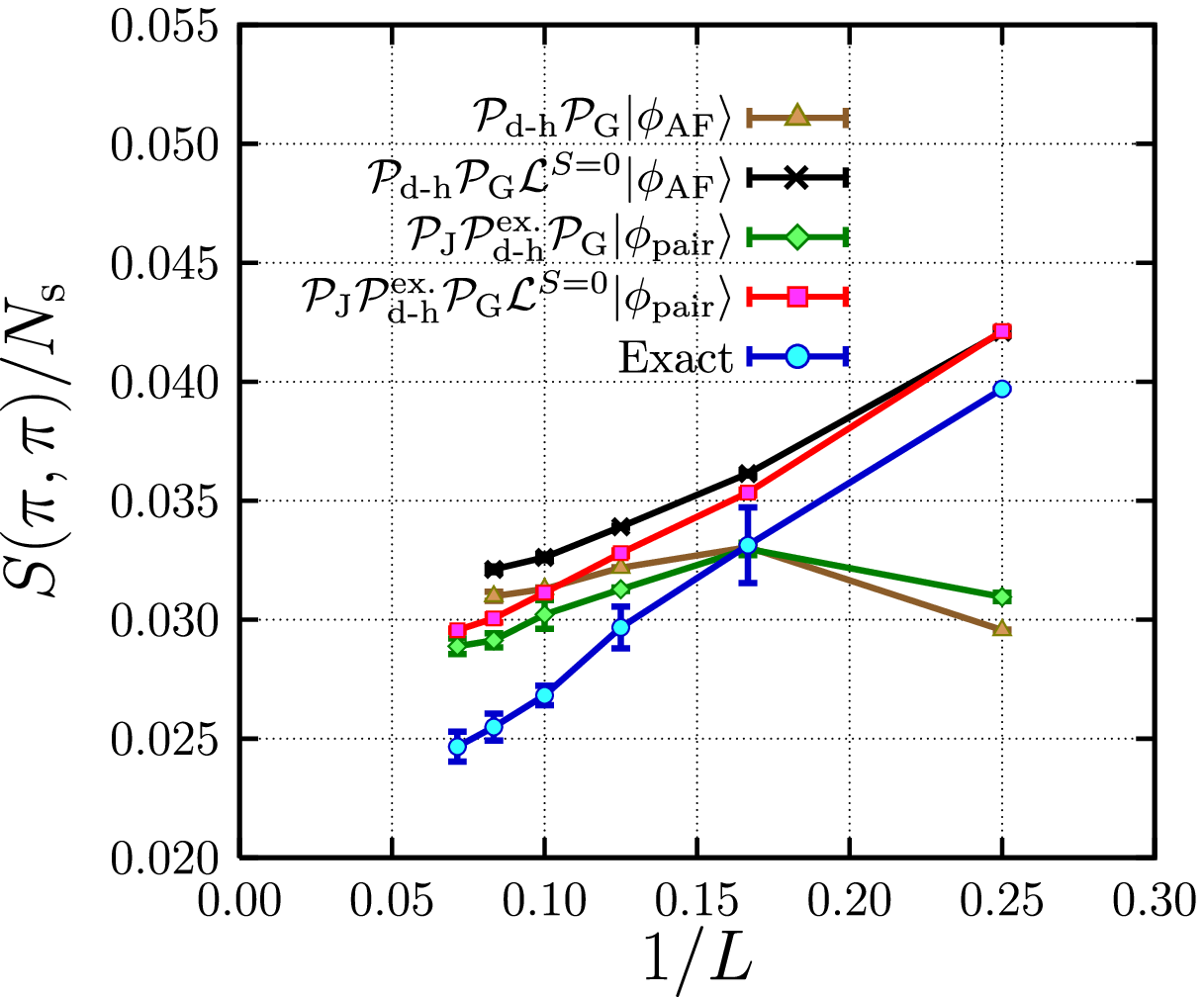}
\caption{%
(Color online) Size dependence of spin structure factor. The exact values are calculated by the same way as Fig. \ref{fig:Res-SizeEn}.
}
\label{fig:Res-SizeSq}
\end{figure}
\par
Figure \ref{fig:Res-SizeDo} shows the size dependence of the double occupancy. $|\phi_{\text{pair}}\ra$ have smaller value than $|\phi_{\text{AF}}\ra$. This difference mainly improves the variational energy. 
Figure \ref{fig:Res-SizeSq} shows the size dependence of spin structure factor 
$S(\pi,\pi)/\Ns$ plotted as a function of $1/L$ following the scaling by the spin-wave theory \cite{Huse}. The behavior of spin-projected wave function is qualitatively better than that of the unprojected case. 
The staggered magnetization in the thermodynamic limit $m_{\text{s}} = [\lim_{N_{\text{s}}\to \infty} S(\pi,\pi)/N_{\text{s}}]^{1/2}$ of each result is estimated as follows:
\[
  m_{\text{s}} \sim \left\{
\begin{array}{@{}r@{\;}c@{\;}l@{}}
  0.173\pm0.004 & : &\text{VMC \ $|\phi_{\text{AF}}\ra$}\\
  0.158\pm0.006 & : &\text{VMC \ $|\phi_{\text{pair}}\ra$}\\
  0.138\pm0.005 & : &\text{AFQMC \cite{WhiteQMC}}\\
  \bigl(\,0.304\pm0.004 & : &\text{Heisenberg model \cite{LiangMag}}\bigr)
\end{array}
  \right.
.
\]
Though the variational results have a general tendency of showing larger $S(\pi,\pi)/\Ns$ and $m_{\text{s}}$ than the AFQMC results, the value obtained from
$|\phi_{\text{pair}}\ra$ shows 
substantial improvement as compared to that obtained from $|\phi_{\text{AF}}\ra$. This means that $|\phi_{\text{pair}}\ra$ treats the AF correlation and quantum fluctuations more correctly than the mean-field descriptions, which often overestimate orders.
\par 
From the above results, the variational wave function constructed from the combination of
$|\phi_{\text{pair}}\ra$, spin projection, and the Gutzwiller-Jastrow factor offers the best description of the ground-state wave function among various choices of variational functions.
\begin{figure}[t]
\centering
\includegraphics[width=0.4\textwidth]{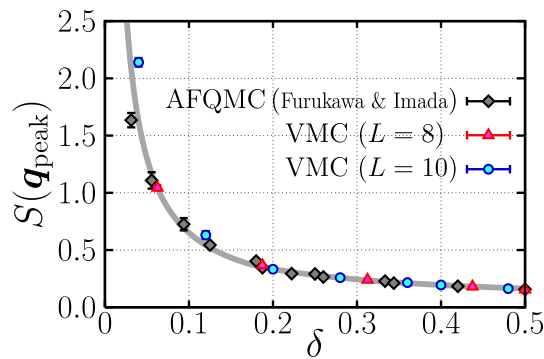}
\caption{%
(Color online) Peak value of spin structure factor $S(\bm{q}_{\text{peak}})$ as a function of the doping concentration $\delta$ for $t'/t=0$ and $U/t=4$. Diamonds are AFQMC results reported in ref. \citen{FurukawaImada}.  Triangles and circles are VMC results obtained by using the variational wave function $\mathcal{P}_{\text{J}}\mathcal{P}_{\text{d-h}}^{\text{ex.}} \mathcal{P}_{\text{G}} \mathcal{L}^{S=0} | \phi_{\text{pair}} \rangle$. Error bars are comparable to the symbol size. The solid curve is the fitting given in Fig. 13(a) of ref. \citen{FurukawaImada}. The fitting satisfies $S(\bm{q}_{\text{peak}}) \propto \delta^{-1}$ for $\delta \le 0.2$.
}
\label{fig:Res-DopeSq}
\end{figure}
%
\subsection{Spin correlation in hole-doped systems}
In order to examine accuracy under severe conditions, we calculate the spin correlation in the hole-doped systems with $t'/t=0$ and $U/t=4$. Figure \ref{fig:Res-DopeSq} shows the doping dependence of the peak value of spin structure factor $S(\bm{q}_{\text{peak}})$. 
Although we are not able to compare our VMC results directly with the AFQMC results in ref. \citen{FurukawaImada} because of the difference in boundary conditions, 
our results show excellent agreement with the unbiased results in a wide range of doping concentration $\delta=1-n$. 
This suggests that quantum fluctuations, especially the short-ranged AF correlation which is crucially important in hole-doped systems, are satisfactorily taken into account in our variational wave function. 
In fact, the system size dependence of $S(\bm{q}_{\text{peak}})$ is small for $\delta \ge 0.05$ in agreement with AFQMC data \cite{FurukawaImada}. 
Our VMC and AFQMC data both consistently show that the AF long range order is restricted to the doping region much smaller than $\delta=0.05$ with the scaling $S(\bm{q}_{\text{peak}}) \propto \delta^{-1}$ in the paramagnetic region ($\delta \ge 0.05$).
%
\subsection{Other results}
We also apply our improved variational wave function to the frustrated Hubbard model with $t'/t=-0.3$ and $n=1$ \cite{TaharaVMCtp0030}. 
The Mott transition between the paramagnetic metal and the AF insulator takes place at $U_{\text{c}}/t=3.3\pm 0.1$ which can be favorably compared with the estimation by the 
PIRG method ($U_{\text{c}}/t \sim 3.6$) \cite{Kashima}. 
This $U_{\text{c}}/t$ is much smaller than the previous variational estimate ($U_{c}/t \sim 6.7$) \cite{YokoVMCdiag}. 
The double occupancy keeps a large value ($D\sim 0.2$) in the metallic phase near the Mott transition and shows very small $U/t$ dependence. In the previous studies, this characteristic feature has been obtained only in the PIRG results. 
The variational wave functions employed in the literature 
include many-body correlations only by much restricted forms, such as the doublon-holon short-ranged factor.
Such restricted form does not sufficiently take into account quantum fluctuations, which are strongly enhanced around the Mott transition. 
Introducing a large number of variational parameters in the Gutzwiller-Jastrow 
factor as well as in the one-body part allows quantitatively accurate treatment of fluctuations with 
complicated correlations. 
\section{Summary and Discussions}
\label{Ch:SummaryDiscussion}
In this paper, we have extended the variational Monte Carlo (VMC) method 
and applied it to the two-dimensional Hubbard model. 
The VMC method originally has several advantages for studies of 
the strongly correlated systems. This method is tractable in large system sizes even 
with strong interactions and geometrical frustrations. 
However, the bias inherently and inevitably contained in the assumed 
variational form of the wave functions is a fundamental drawback in the VMC method.
\par
In order to overcome and go beyond the conventional limitation in the VMC framework, 
we have improved variational wave functions by the following extensions:
\begin{enumerate}[(i)]
\item By introducing a large number of variational parameters, we have constructed the one-body part including various states such as paramagnetic metals, antiferromagnetically ordered states, and superconducting states with any wavenumber (spatial) dependence of gap functions within a single functional form. 
Moreover, $|\phi_{\text{pair}}\ra$ enables efficient treatment of quantum fluctuations of spins. 
This extension efficiently reduces biases coming from assumed states in the previous VMC method.
\item We have introduced a new factor 
for the variational wave 
functions: 
The quantum-number projection factor restores the inherent symmetry of the wave function and, as a result, the accuracy is substantially 
improved.
\item We have combined our improvements with the recently improved Gutzwiller-Jastrow factor including many variational parameters. In particular, the $\vk$-dependent Jastrow factor efficiently takes into account fluctuations of charge.
\end{enumerate}
\par
The accuracy of our variational framework has been examined by the comparison with the unbiased results obtained from the exact diagonalization and the auxiliary-field quantum Monte Carlo method. 
It has turned out that the improvement of the one-body part and 
the quantum-number projection enable 
highly accurate descriptions of the wave function for 
the ground state. 
In the system with 
$t'/t=0$, $U/t=4$, and $n=1$, the relative error 
reaches as low as 
$0.5\%$ at $L=4$ and $1\%$ in the thermodynamic limit.
These errors are typically a half of the best available results in the literature \cite{Baeriswyl}.
\par
Our improvement of the variational wave function does not change the basic numerical framework of the VMC method, namely sampling of the real space configurations. Therefore, our approach is able to combine the ``post-VMC'' method such as the Lanczos method \cite{Heeb} and the diffusion Monte Carlo method \cite{DMC}. 
These additional treatments certainly even more reduce the biases. 
In this paper, we have used a single pairing wave function for the core one-body part. The linear combination of several pairing wave functions for the core will improve the variational wave function, though this causes linear increase of the computational costs. 
The search for efficient representations of this multi-configuration is a future problem.
\par
Our improvement of variational Monte Carlo method opens a possibility of studying strongly correlated electron systems by reducing the effects of biases from restricted variational forms. In particular, effects of short-ranged spin and charge fluctuations may be studied with quantitative accuracy. Applications of our refined algorithms will be reported elsewhere \cite{TaharaVMCtp0030}.
\section*{Acknowledgments}
One of the authors (D.T.) thanks 
 S.~Watanabe and T.~Misawa for useful discussions. This work was supported by Grants-in-Aid for
Scientific Research on Priority Areas under the grant numbers 
17071003, 16076212, and 17064004 
from the Ministry of Education, Culture, Sports, Science and
Technology. A part of our computation has been done using
the facilities of the Supercomputer Center, Institute for Solid
State Physics, University of Tokyo.
%
\appendix
\section{Relation between $\sP_{\text{G}}^{\infty} | \phi_{\text{pair}}\ra$ and the RVB basis}
\label{App:RVB}
The wave functions based on the resonating valence bond (RVB) basis \cite{LDA} provide highly accurate 
descriptions in spin systems. 
As discussed by Liang, 
Doucot, and Anderson \cite{LDA}, these functions can control AF correlations. 
It is worth 
clarifying the relation between the RVB wave function and wave functions used in itinerant electron systems.
As pointed out first by Anderson \cite{AndersonPBCS}, 
the RVB wave function is equivalent to the so called ``projected BCS'' wave function.
\par
\begin{figure}[htbp]
\centering
\includegraphics[width=0.35\textwidth]{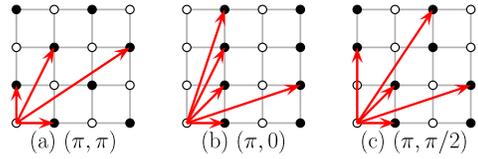}
\caption{%
(Color online) Possible $A$-$B$ sublattice pattern of (a) $(\pi,\pi)$-order, (b) $(\pi,0)$-order, and (c) $(\pi,\pi/2)$-order. 
Open circles and filled circles are $A$ and $B$ sublattice respectively.
Vectors are some examples corresponding to nonvanishing elements $f(\vr)$.
}
\label{fig:VWF-RVB}
\end{figure}

The RVB wave function is constructed by singlet-dimer covering as
\[
  |\phi_{\text{RVB}}\ra = \sum_{\{C\}} \Biggl[ \prod_{k=1}^{N/2} h_{i_kj_k}
    \Bigl( 
        c_{i_k\uparrow}^\dag c_{j_k\downarrow}^\dag
      - c_{i_k\downarrow}^\dag c_{j_k\uparrow}^\dag
    \Bigr)
  \Biggr] |0\ra
,
\]
where the summation $\sum_{\{C\}}$ runs over all the 
patterns of coverings $\{C\}$ with some conditions described below. 
When the AF order has $A$ and $B$ sublattices, 
two conditions ``$i_k\in A$'' and ``$j_k\in B$'' are imposed on 
$\{C\}$. 
Since two singlet dimers sharing one same site make a doublon:
$    ( 
        c_{i\uparrow}^\dag c_{j\downarrow}^\dag
      - c_{i\downarrow}^\dag c_{j\uparrow}^\dag
    )
    ( 
        c_{i\uparrow}^\dag c_{k\downarrow}^\dag
      - c_{i\downarrow}^\dag c_{k\uparrow}^\dag
    )
=
    ( 
        c_{j\uparrow}^\dag c_{k\downarrow}^\dag
      - c_{j\downarrow}^\dag c_{k\uparrow}^\dag
    )
        c_{i\uparrow}^\dag c_{i\downarrow}^\dag
$ \, $
  (i\neq j, \, j\neq k, \, k\neq i)
$,
$|\phi_{\text{RVB}}\ra$ can be rewritten as
\begin{align} \notag
  |\phi_{\text{RVB}}\ra &= \sP_{\text{G}}^{\infty}
  \Biggl[
    \sum_{(i,j):\{C\}} h_{ij} 
    \Bigl( 
        c_{i\uparrow}^\dag c_{j\downarrow}^\dag
      - c_{i\downarrow}^\dag c_{j\uparrow}^\dag
    \Bigr)
  \Biggr]^{N/2} |0\ra
\\&=
  \sP_{\text{G}}^{\infty} 
  \Biggl[
    \sum_{i,j=1}^{\Ns} f_{ij} 
        c_{i\uparrow}^\dag c_{j\downarrow}^\dag
  \Biggr]^{N/2} |0\ra
,
\end{align}
where $\sP_{\text{G}}^{\infty} = \prod_i (1-n_{i\uparrow}n_{i\downarrow})$ is the Gutzwiller factor (eq. (\ref{eq:VWF-GutzInf})) which filters out all the components with a finite double 
occupancy, $\sum_{(i,j):\{C\}}$ sums up all the 
$(i,j)$ pairs appearing in $\{C\}$, and off-diagonal elements of $f_{ij}$ are related to $h_{ij}$ as
\[ \label{eq:VWF_RVBpair}
  f_{ij}=f_{ji} = \left\{
\begin{array}{@{\,}c@{\ }l@{}}
  h_{ij} & \text{($i\in A$, $j\in B$) or ($i\in B$, $j\in A$)} \\
  0      & \text{($i\in A$, $j\in A$) or ($i\in B$, $j\in B$)}
\end{array}
  \right.
.
\]
The diagonal elements $f_{ii}$ are arbitrary because of $\sP_{\text{G}}^{\infty}$. 
Equation (\ref{eq:VWF_RVBpair}) means that $f_{ij}=f(\vr_i-\vr_j)$ depends on the relative vector ($\vr_i-\vr_j$) and the translational symmetry is preserved. 
Figure (\ref{fig:VWF-RVB}) shows three examples of $A$-$B$ sublattice patterns and some relative vectors corresponding to nonvanishing elements $f(\vr)$.
By using the Fourier transformation, $|\phi_{\text{RVB}}\ra$ is written as
\[
  |\phi_{\text{RVB}}\ra = 
  \sP_{\text{G}}^{\infty} 
  \Biggl[
    \sum_{\vk} f_{\vk} 
        c_{\vk\uparrow}^\dag c_{-\vk\downarrow}^\dag
  \Biggr]^{N/2} |0\ra
=
  |\text{p-BCS}\ra
\]
with 
\[
  f_{\vk} = \sum_{\vr} f(\vr) e^{-i\vk\bdot\vr}
.
\]
Since $[\sum_{\vk}f_{\vk}c_{\vk\uparrow}^\dag c_{-\vk\downarrow}^\dag ]^{N/2}|0\ra$ is the BCS wave function, this means that the RVB wave function corresponds to the BCS wave function with $\sP_{\text{G}}^{\infty}$, which is called a ``projected BCS'' wave function $|\text{p-BCS}\ra$.
\par
From the above discussion, the variational wave function constructed from $|\phi_{\text{pair}}\ra$ and $\sP_{\text{G}}$ includes the RVB basis. 
Determination of $f_{ij}$ or $h_{ij}$ by hand is discussed in detail in the literature \cite{RVBweight}. 
In our calculation, $f_{ij}$ is numerically determined by the optimization of variational parameter $\varphi^{(1)}(\vk)$ in eq. (\ref{eq:VWF-one-bodyfull}).
%
\section{Linear Algebra of Skew-symmetric Matrix and Pfaffian}
\label{App:PfMat}
In this appendix, some useful relations between the skew-symmetric matrices and the 
Pfaffians are derived 
\cite{Bajdich}. We assume that all the 
elements of the 
matrices are real numbers.
\subsection{Definition}
A $2M\times 2M$ skew-symmetric matrix $\mA$ satisfies the following relation:
\[
  \mA^{T} = -\mA \quad \bigl( A_{ij} = -A_{ji}\bigr),
\]
where $\mA^T$ denotes the transposed matrix of $\mA$. The Pfaffian of $\mA$ is defined as antisymmetrized product
\begin{align} \notag
  \Pf A &= \mathcal{A} [A_{12}A_{34} \cdots A_{2M-1,2M}]
\\&=
  \sum_{\mathcal{P}} (-1)^{\mathcal{P}} \prod_{k=1}^{M} A_{\sP(2k-1)\sP(2k)},
\end{align}
where the sum runs over all the pair 
partitions $\mathcal{P}$ of $2M$ indices 
such that $\sP(2k-1) < \sP(2k)$. 
Here, $(-1)^{\mathcal{P}}$ is the parity of the permutation $\sP$:
\[
    \left(
      \begin{array}{@{}ccccc@{}}
        1   & 2   & \cdots & 2M-1  & 2M \\
        \sP(1) & \sP(2) & \cdots & \sP(2M-1) & \sP(2M)
      \end{array}
    \right).
\]
\subsection{Identities}
\subsubsection{Basic relations}
\allowdisplaybreaks
The Pfaffian satisfies the following relations:
\begin{align}
\Pf \mA^T &= (-1)^M \Pf \mA
\\
(\Pf \mA)^2 &= \det \mA
\\
\Pf \left[
\begin{array}{@{\,}c@{\:}c@{\,}}
  \mA_1 & 0 \\ 0 & \mA_2
\end{array}
\right] &= \Pf \mA_1 \times \Pf \mA_2
\\
\Pf (\mB\mA\mB^T) &= \det \mB \times \Pf \mA
\\
\Pf \left[
\begin{array}{@{\,}c@{\:}c@{\,}}
  0 & \mC \\ -\mC^T & 0
\end{array}
\right] &= (-1)^{M(M-1)/2} \det \mC
\end{align}
where $\mA$, $\mA_1$, $\mA_2$ are $2M\times 2M$ skew-symmetric matrices, $\mB(2M\times 2M)$ and $\mC(M\times M)$ are arbitrary real matrices.
\par
By taking $\mB$ as elementary transformation matrices, we can verify the following properties of Pfaffians:
\begin{enumerate}[(i)]
\item Multiplication of a row and corresponding column by a constant is equivalent to multiplication of original Pfaffian by the same constant.
\item Interchange of two different rows and corresponding columns changes the sign of Pfaffian.
\item Consider a vector $\bm{p}$ which is the same as a row of $\mA$
and another transposed vector $\bm{q}^{T}$ which is the same
as the corresponding column of $\mA$.
Addtion of $\bm{p}$ 
to another row and 
addition of $\bm{q}^{T}$ to the corresponding column does not change 
the value of Pfaffian. 
\end{enumerate}
By performing a Gaussian elimination technique and pivoting rows and corresponding columns, we can transform any skew-symmetric matrix into block-diagonal form and obtain the value of Pfaffian:
\begin{align}\notag 
  \Pf \mA &= (-1)^{p} \Pf \left[
\begin{array}{@{\,}c@{\:}c@{\:}c@{\:}c@{\:}c@{\:}c@{\:}c@{\,}}
  0 & \lambda_1 & & & & & \\
  -\lambda_1 &0 & & & & 0& \\
  & & 0 & \lambda_2 & & & \\
  & & -\lambda_2 &0 & & & \\
  & & & & \ddots & & \\
  & 0& & & & 0 & \lambda_M \\
  & & & & & -\lambda_M &0 \\
\end{array}
  \right] 
\\&= (-1)^p \lambda_1 \lambda_2 \cdots \lambda_M,
\end{align}
where $p$ is the frequency of pivoting and $\lambda_i$ ($i=1,\cdots,M$) are results of the elimination. 
The calculation of a Pfaffian costs $\mathcal{O}(M^3)$ operations.
\subsubsection{Cayley's identity}
Cayley showed a useful identity \cite{Cayley}:
\begin{align}\notag\tiny
\det \left[\begin{array}{@{\,}c@{\:}c@{\:}c@{\:}c@{\,}}
0  & A_{12}  &\ldots &  A_{1M}\\
b_{12}  & 0  & \ldots &  A_{2M}\\
 \vdots & \vdots &  \ddots &  \vdots \\
b_{1M} & -A_{2M} & \ldots &  0\\
\end{array}\right]
=&\tiny
\Pf \left[\begin{array}{@{\,}c@{\:}c@{\:}c@{\:}c@{\,}}
0  & A_{12}  &\ldots &  A_{1M}\\
-A_{12}  & 0 & \ldots &  A_{2M}\\
 \vdots &\vdots &  \ddots &  \vdots \\
-A_{1M} & -A_{2M} & \ldots &  0\\
\end{array}\right]
\\
&\!\!\!\!\!\!\!\tiny\times
\Pf \left[\begin{array}{@{\,}c@{\:}c@{\:}c@{\:}c@{\,}}
0  & -b_{12}  &\ldots &  -b_{1M}\\
b_{12}  & 0  & \ldots &  A_{2M}\\
 \vdots & \vdots &  \ddots &  \vdots \\
b_{1M} & -A_{2M} & \ldots &  0\\
\end{array}\right].
\end{align}
From this identity and the cofactor expansion of determinant, we can obtain the relation between the Pfaffian and the inverse matrix of a skew-symmetric matrix $\mA$ and the Pfaffian of a skew-symmetric matrix $\mB$ which has same elements of $\mA$ except for $\alpha$-th row and column:
\[ \label{eq:App.Pf.Upadate01}
  \Pf \mB = \frac{\det \mB}{\Pf \mA}
=
  \frac{\det \mA \sum_{m} A_{\alpha m}^{-1}b_m}{\Pf \mA}
=  \Pf \mA \sum_{m} A_{\alpha m}^{-1}b_m
,
\]
where $b_m$ ($m=1,\cdots,M$) are the updated elements of $\alpha$-th row in $\mB$. 
eq. (\ref{eq:App.Pf.Upadate01}) allows us to calculate $\Pf \mB$ with $\mathcal{O}(M)$ operations from $\Pf \mA$ and $\mA^{-1}$.
\subsection{Update technique for skew-symmetric matrix}
We derive one of the most important techniques for VMC with Pfaffians, which is similar to the update technique for VMC with determinants \cite{CeperleyVMC}.
\subsubsection{Preparation}
For any nonsingular matrix $\mA$ and any column vector $\bm{u}$ and $\bm{v}$ with the condition $1+\bm{v}^T \mA \bm{u}\neq 0$, we have the Sherman-Morrison's formula \cite{Sherman-Morrison}:
\begin{gather}
  \bigl[ \mA + \bm{u} \bm{v}^T\bigr]_{ij}^{-1} 
=
  A_{ij}^{-1} - \frac{1}{1+\bm{v}^T \mA^{-1} \bm{u}} \sum_{m,n}A_{im}^{-1} u_m v_n A_{nj}^{-1}.
\end{gather}
If we take $\mA$ as a skew-symmetric matrix ($\mA^T=-\mA$), then we can derive an inverse matrix of $\mB=\mA+\bm{u}\bm{v}^T-\bm{v}\bm{u}^T$;
\begin{align}
\notag
  B_{ij}^{-1} = A_{ij}^{-1} + \frac{1}{1+\bm{v}^T \mA^{-1} \bm{u}}
  \sum_{m,n}\Bigl[
    &A_{im}^{-1}v_m  u_n A_{nj}^{-1}
\\\label{eq:App.SM01}
  &- A_{im}^{-1}u_m  v_n A_{nj}^{-1}
  \Bigr].
\end{align}
\subsubsection{Update formula for inverse matrix}
An inverse matrix of $\mB$ with updated $\alpha$-th row and column from the original skew-symmetric matrix $\mA$ are calculated by
\begin{align}
\notag
  B_{ij}^{-1} =& A_{ij}^{-1} + \frac{1}{\sum_m A_{\alpha m}^{-1} b_m}
  \Biggl\{
  - \biggl[ \sum_m A_{im}^{-1} b_m \biggr] A_{\alpha j}^{-1}
\\ \label{eq:App.Inv.Upadate01}
&
  + \biggl[ \sum_m A_{jm}^{-1} b_m \biggr] A_{\alpha i}^{-1}
  + \delta_{i\alpha} A_{\alpha j}^{-1}
  - \delta_{j\alpha} A_{\alpha i}^{-1}
  \Biggr\},
\end{align}
where $b_m$ ($m=1,\cdots,M$) are the updated elements of $\alpha$-th row in $\mB$ and $\delta_{ij}$ is the Kronecker's delta.
If $\mB$ is a singular matrix ($\Pf \mB=0$), the following formula 
holds instead of eq. (\ref{eq:App.Inv.Upadate01}):
\begin{align}\notag
  \bigl[ \Pf \mB \cdot \mB^{-1}\bigr]_{ij} = &\Pf \mA \Biggl\{
    \biggl[\sum_{m} A_{\alpha m}^{-1} b_m \biggr] A_{ij}^{-1}
\\\notag
&
  - \biggl[ \sum_m A_{im}^{-1} b_m \biggr] A_{\alpha j}^{-1}
  + \biggl[ \sum_m A_{jm}^{-1} b_m \biggr] A_{\alpha i}^{-1}
\\ \label{eq:App.Inv.Upadate02}
&
  + \delta_{i\alpha} A_{\alpha j}^{-1}
  - \delta_{j\alpha} A_{\alpha i}^{-1}
  \Biggr\},
\end{align}
where $[\Pf \mB\cdot \mB^{-1}]$ is a symbolic notation. 
Above formulas can be derived by using eq. (\ref{eq:App.SM01}) and by 
taking
\[
\left\{
\begin{array}{l}
  u_i = b_i - A_{\alpha i} \\
  v_i = \delta_{\alpha i}
\end{array}
\right. .
\]
Equations (\ref{eq:App.Inv.Upadate01}) and (\ref{eq:App.Inv.Upadate02}) allows us to calculate $\mB^{-1}$ 
and $[\Pf \mB\cdot \mB^{-1}]$ with $\mathcal{O}(M^2)$ operations 
from $\Pf \mA$ and $\mA^{-1}$, 
which may be compared with $\mathcal{O}(M^3)$ operation if one calculates
from scratch.


\begin{thebibliography}{99}

\bibitem{ImadaFujimoriTokura} 
For a review see M.~Imada, A.~Fujimori, and Y.~Tokura: \journal{\RMP}{70}{1039}{1998}.

\bibitem{Mott}
N.~F.~Mott and R.~Peierls: \journal{Proc.\ Phys.\ Soc.\ London}{49}{72}{1937}.

\bibitem{Bednorz} 
J.~G.~Bednorz and K.~A.~M\"{u}ller: \journal{Z.\ Phys.\ B}{64}{189}{1986}.

\bibitem{QMC01}
R.~Blankenbecler, D.~J.~Scalapino, and R.~L.~Sugar: \journal{\PRD}{24}{2278}{1981}.

\bibitem{QMC02}
S.~Sorella, S.~Baroni, R.~Car, and M.~Parrinello: \journal{Europhys.\ Lett.}{8}{663}{1989}.

\bibitem{AFQMC}
M.~Imada and Y.~Hatsugai: \journal{\JPSJ}{58}{3752}{1989}.

\bibitem{FurukawaImada}
N.~Furukawa and M.~Imada: \journal{\JPSJ}{61}{3331}{1992}.

\bibitem{DMRG}
S.~R.~White: \journal{\PRB}{48}{10345}{1993}.

\bibitem{DMFT01}
W.~Metzner and D.~Vollhardt: \journal{\PRL}{62}{324}{1989}.

\bibitem{DMFT02}
A.~Georges, G.~Kotliar, W.~Krauth, and M.~J.~Rozenberg: \journal{\RMP}{68}{13}{1996}.

\bibitem{PIRGx01}
M.~Imada and T.~Kashima: \journal{\JPSJ}{69}{2723}{2000}.

\bibitem{PIRGx02}
T.~Kashima and M.~Imada: \journal{\JPSJ}{70}{2287}{2001}.

\bibitem{Kashima}
T.~Kashima and M.~Imada: \journal{\JPSJ}{70}{3052}{2001}.

\bibitem{Morita}
H.~Morita, S.~Watanabe, and M.~Imada: \journal{\JPSJ}{71}{2109}{2002}.

\bibitem{GPIRG}
S.~Watanabe and M.~Imada: \journal{\JPSJ}{73}{1251}{2004}.

\bibitem{PIRGMizusaki}
T.~Mizusaki and M.~Imada: \journal{\PRB}{69}{125110}{2004}.

\bibitem{GBMC01}
J.~F.~Corney and P.~D.~Drummond: Phys.\ Rev.\ Lett.\ \textbf{93}, 260401 (2004); 
Phys.\ Rev.\ B \textbf{73}, 125112 (2006); J.\ Phys.\ A: Math.\ Gen.\ \textbf{39}, 269 (2006).

\bibitem{GBMCAssaad}
F.~F.~Assaad, P.~Werner, P.~Corboz, E.~Gull, and M.~Troyer: \journal{\PRB}{72}{224518}{2005}.

\bibitem{GBMCAimi}
T.~Aimi and M.~Imada: \journal{\JPSJ}{76}{084709}{2007}.

\bibitem{CeperleyVMC} 
D.~Ceperley, G.~V.~Chester, and M.~H.~Kalos: \journal{\PRB}{16}{3081}{1977}.

\bibitem{Jastrow}
R.~Jastrow: \journal{\PR}{98}{1479}{1955}.

\bibitem{Gutzwiller}
M.~C.~Gutzwiller: \journal{\PRL}{10}{159}{1963}.

\bibitem{SorellaSR}
S.~Sorella: \journal{\PRB}{64}{024512}{2001}.

\bibitem{SorellaSRH}
S.~Sorella: \journal{\PRB}{71}{241103}{2005}.

\bibitem{VMCOptLM02}
C.~J.~Umrigar, J.~Toulouse, C.~Filippi, S.~Sorella, and R.~G.~Hennig: \journal{\PRL}{98}{110201}{2007}.

\bibitem{VMCOptNW01}
X.~Lin, H.~Zhang, and A.~M.~Rappe: \journal{\JCP}{112}{2650}{2000}.

\bibitem{VMCOptLM01}
M.~P.~Nightingale and V.~Melik-Alaverdian: \journal{\PRL}{87}{043401}{2001}.

\bibitem{VMCOptNW02}
C.~J.~Umrigar and C.~Filippi: \journal{\PRL}{94}{150201}{2005}.

\bibitem{VMCOptEFP}
A.~Scemama and C.~Filippi: \journal{\PRB}{73}{241101}{2006}.

\bibitem{CapelloPRL01}
M.~Capello, F.~Becca, M.~Fabrizio, S.~Sorella, and E.~Tosatti: \journal{\PRL}{94}{026406}{2005}.

\bibitem{ManyBody}
P.~Ring and P.~Schuck: {\it The Nuclear Many-Body Problem}, (Springer-Verlag, New York, Heidelberg, Berlin, 1980).

\bibitem{Giamarchi}
T.~Giamarchi and C.~Lhuillier: \journal{\PRB}{43}{12943}{1991}.

\bibitem{HimedaOgata}
A.~Himeda and M.~Ogata: \journal{\PRL}{85}{4345}{2000}.

\bibitem{LDA}
S.~Liang, B.~Doucot, and P.~W.~Anderson: \journal{\PRL}{61}{365}{1988}.

\bibitem{CasulaJCP} 
M.~Casula, C.~Attaccalite, and S.~Sorella: \journal{\JCP}{121}{7110}{2004}.

\bibitem{Kaplan}
T.~A.~Kaplan, P.~Horsch, and P.~Fulde: \journal{\PRL}{49}{889}{1982}.

\bibitem{YokoyamaShiba3}
H.~Yokoyama and H.~Shiba: \journal{\JPSJ}{59}{3669}{1990}.

\bibitem{Liu}
J.~Liu, J.~Schmalian, and N.~Trivedi: \journal{\PRL}{94}{127003}{2005}.

\bibitem{WataVMC01}
T.~Watanabe, H.~Yokoyama, Y.~Tanaka, and J.~Inoue: \journal{\JPSJ}{75}{074707}{2006}.

\bibitem{YokoVMCdiag}
H.~Yokoyama, M.~Ogata, and Y.~Tanaka: \journal{\JPSJ}{75}{114706}{2006}.

\bibitem{KobayashiYokoyama} 
K.~Kobayashi and H.~Yokoyama: Physica C, \textbf{463}-\textbf{465} (2007) 141-145.

\bibitem{NumRec}
W.~H.~Press, S.~A.~Teukolsky, W.~T.~Vetterling, and B.~P.~Flannery: \textit{NUMERICAL RECIPES in C} (Cambridge University Press, 1993).

\bibitem{PfVMC01}
J.~P.~Bouchaud, A.~Georges, and C.~Lhuillier: \journal{J.\ Phys. (Paris)}{49}{553}{1988}.

\bibitem{SorellaJCP}
S.~Sorella, M.~Casula, and D.~Rocca: \journal{\JCP}{127}{014105}{2007}.

\bibitem{Huse}
D.~A.~Huse: \journal{\PRB}{37}{2380}{1988}.

\bibitem{WhiteQMC}
By using the extrapolated value in the early AFQMC study (%
S.~R.~White, D.~J.~Scalapino, R.~L.~Sugar, E.~Y.~Loh, J.~E.~Gubernatis, and R.~T.~Scalettar: \journal{\PRB}{40}{506}{1989}%
), the staggered magnetization $m_{\text{s}}$ is estimated to be about $0.10$. 
However, we have reexamined the system size dependence by using our AFQMC calculations. 
It has turned out that the system size ($L \le 10$) is not large enough for the extrapolation. 
The extrapolation with larger sizes up to $L=14$ as illustrated in Fig. \ref{fig:Res-SizeSq} gives $m_{\text{s}} \sim 0.14$ irrespective of the boundary conditions.

\bibitem{LiangMag}
S.~Liang: \journal{\PRB}{42}{6555}{1990}.

\bibitem{TaharaVMCtp0030}
D.~Tahara and M.~Imada: \journal{\JPSJ}{77}{093703}{2008}.

\bibitem{Baeriswyl}
We note that the variational wave function proposed by Baeriswyl %
(D.~Baeriswyl: \textit{Nonlinearity on Condensed Matter}, edited by A.~R.~Bishop, \etal, Springer Series in Solis State Sciences Vol. \text{69} (Springer, Berlin, 1987), p.~183; D.~Eichenberger and D.~Baeriswyl: \journal{\PRB}{76}{180504}{2007}) 
is more accurate than our results in small $U/t$ and small cluster systems. The relative error of energy is about $0.25$\% for $U/t=4$, $\Ns=10$, and $n=1$ (H.~Otsuka, \journal{\JPSJ}{61}{1645}{1991}). 
However, the errors of this wave function are rapidly enhanced when $U/t$ or $\Ns$ increases. The errors in the above paper are about $13$\% and $2$\% for $(U/t,\Ns,n)=(20,10,1)$ and $(4,64,1)$ systems, respectively. On the other hand, our results are $2$\% and $0.8$\% for $(U/t,\Ns,n)=(20,16,1)$ and $(4,64,1)$ systems, respectively. 
It is difficult to improve systematically by introducing additional Gutzwiller-Jastrow factors, because the MC sampling of the former is based on the Stratonovich-Hubbard transformation. 
Our improvements offer accurate variational wave functions even in systems with large $U/t$ and/or geometrical frustration effects.

\bibitem{Heeb}
E.~S.~Heeb and T.~M.~Rice: \journal{Europhys.\ Lett.}{27}{673}{1994}.

\bibitem{DMC}
D.~M.~Ceperley and B.~J.~Alder: \journal{\PRL}{45}{566}{1980}: \journal{Science}{231}{555}{1986}.

\bibitem{AndersonPBCS}
P.~W.~Anderson: \journal{Science}{235}{1196}{1987}.

\bibitem{RVBweight}
For example, N.~Read and B.~Chakraborty: \journal{\PRB}{40}{7133}{1989}; S.~Yunoki and S.~Sorella: \journal{\PRB}{74}{014408}{2006}.

\bibitem{Bajdich}
Algebra and formulae are collected in M.~Bajdich, L.~Mitas, L.~K.~Wagner, and K.~E.~Schmidt: \journal{\PRB}{77}{115112}{2008}.

\bibitem{Cayley}
A.~Cayley: {\sl Sur les d\' eterminants gauches}, 
J. reine angew. Math. {\bf 38}, pp. 93-96 (1849); reprinted in 
{\sl The collected mathematical papers of Arthur Cayley}, Cambridge [Eng.] The University Press, Cambridge, vol. 1, pp. 410-413 (1889).

\bibitem{Sherman-Morrison}
J.~Sherman and W.~J.~Morrison: \journal{Ann.\ Math.\ Stat.}{20}{621}{1949}.
\end{thebibliography}
\end{document}